\documentclass[preprint,12pt]{elsarticle}

\usepackage{amsmath,amsfonts,amssymb,amsthm,booktabs,color,epsfig,graphicx,hyperref,url}
\usepackage{amsmath, amssymb, enumerate, amsthm}
\usepackage{lineno,hyperref}
\usepackage{amsfonts}
\usepackage{enumerate}
\usepackage{graphicx,psfrag,color}
\usepackage{amsfonts}
\usepackage{enumerate}
\usepackage{threeparttable}
\usepackage{algorithm,algcompatible,amsmath}
\algnewcommand\INPUT{\item[\textbf{Input:}]}%
\algnewcommand\OUTPUT{\item[\textbf{Output:}]}%
\usepackage{CJK}
\usepackage{graphicx,psfrag,color}
\usepackage{fancyhdr}
\usepackage{amsmath,amsthm,amssymb}
\usepackage{mathrsfs}
\usepackage{multirow}
\usepackage{color}
\usepackage{indentfirst}
\usepackage{epsfig, epsfig, graphics, lscape, amsbsy, amstext, natbib}
\usepackage{graphicx}
\usepackage{subcaption}
\usepackage{algorithm}
\usepackage{multirow}
\usepackage{array}
\usepackage[noend]{algpseudocode}
\usepackage{tikz}
\usepackage{pgfplots}
\newtheorem{theorem}{THEOREM}

\newtheorem{lemma}{LEMMA}

\newcommand{\T}{\!\top\!}
\usepackage{algorithm,algpseudocode}
\algnewcommand{\Inputs}[1]{%
  \State \textbf{Inputs:}
  \Statex \hspace*{\algorithmicindent}\parbox[t]{.8\linewidth}{\raggedright #1}
}
\algnewcommand{\Initialize}[1]{%
  \State \textbf{Initialize:}
  \Statex \hspace*{\algorithmicindent}\parbox[t]{.8\linewidth}{\raggedright #1}
}

\modulolinenumbers[5]

\begin{document}

\begin{frontmatter}

\title{Optimal model averaging forecasting in high-dimensional
survival analysis}

\author{Xiaodong Yan, Hongni Wang, Wei Wang, Jinhan Xie*, Yanyan Ren*, Xinjun Wang*}

\begin{abstract}
This article considers ultrahigh-dimensional forecasting problems with survival response
variables. We propose a two-step model averaging procedure for improving the forecasting accuracy of the true conditional mean of a survival response variable. The first step is
to construct a class of candidate models, each with low-dimensional covariates. For this,
a feature screening procedure is developed to separate the active and inactive predictors
through a marginal Buckley¨CJames index, and to group covariates with a similar index
size together to form regression models with survival response variables. The proposed
screening method can select active predictors under covariate-dependent censoring, and
enjoys sure screening consistency under mild regularity conditions. The second step is to
find the optimal model weights for averaging by adapting a delete-one cross-validation
criterion, without the standard constraint that the weights sum to one. The theoretical
results show that the delete-one cross-validation criterion achieves the lowest possible
forecasting loss asymptotically. Numerical studies demonstrate the superior performance
of the proposed variable screening and model averaging procedures over existing
methods.
\end{abstract}

\end{frontmatter}

\section{Introduction}\label{section1}
Improving forecasting accuracy for the true conditional mean of survival response variable with ultrahigh-dimensional predictors  is a challenging problem in many scientific fields such as economics, finance (Stepanova and Thomas 2002; Dirick et al. 2017, public health, genomics and medicine (Tang et al. 2020; Yan et al. 2020).  Examples include gene expression data influencing the health of people such as the mantle cell lymphoma data that motivated this research. It has long been recognized that  model selection and model averaging are two popular methods 
 for enhancing forecasting accuracy in
regression analysis (Burnham and Anderson 2002; Yuan
and Yang 2005).
Model selection methods have been used
to determine a `correct' model when there are several candidate models to choose from but
no definite scientific rationale to dictate which one should be used. Different model selection methods or criteria may yield different best models. Hence, inference based on the final model can be seriously misleading (Hjort and
Claeskens, 2003). Instead of relying on only one model, model averaging compromises across
a set of competing models by assigning different weights. In doing so, model uncertainty
is incorporated into the conclusions about unknown parameters. Therefore, predictions
obtained by model averaging are often robust in the sense that they reduce model selection
bias and account for model selection uncertainty (Zhang et al. 2014, 2016; Xie et al. 2020).

 For the analysis of high-dimensional data, many penalized methods, such as the least absolute shrinkage and selection operator (Lasso; Tibshirani 1996), the smoothly
clipped absolute deviation (SCAD; Fan and Li 2001), the adaptive Lasso (Zou 2006), the Dantzig
selector (Candes and Tao 2007), and the minimax concave penalty (MCP; Zhang 2010) have been developed to simultaneously select the important covariates and estimate parameters in various statistical
models when the number of covariates diverges.
High-dimensional sparse modeling with survival data is of great practical importance. 
It is commonly assumed that only a small number of covariates actually contributes to the considered survival models, which leads to the well-known sparse survival models for helping interpretation and improving forecasting accuracy (Bradic, Fan, and Wang 2011).
Several regularization methods originally developed for linear regression have been adapted to survival models. For example, Tibshirani (1997) and Fan and Li (2002) extended the Lasso and nonconcave penalized likelihood, respectively, to the Cox model, while Zhang and Lu (2007) and Zou (2008) developed the adaptive Lasso and the efficient and adaptive shrinkage methods for variable selection in the Cox model. Antoniadis, Fryzlewicz, and Letu$\acute{e}$ (2010) studied the Dantzig selector for the Cox model in a high-dimensional setting. Variable selection techniques  have also been extended to other survival models including the additive hazards model (Leng and Ma 2007; Martinussen and Scheike 2009; Lin and Lv 2013) and the accelerate failure time model (Huang et al. 2006;  Wang et al. 2008).

On the other hand, a number of model averaging procedures are developed for unsurvival response variables  under the standard setting in which the number of predictors is   fixed and much smaller than the sample size.
These methods include the forecasting model averaging  (Newbold and Granger
1974), the least squares  model averaging  (Hansen 2007; Wan, Zhang, and Zou
2010), the predictive
likelihood-based model averaging (Ando and Tsay 2010), the frequentist model averaging (Liang
et al. 2011), the jackknife model averaging (Hansen and Racine 2012), the  heteroskedasticity-robust $C_p$ model averaging
(Liu and Okui 2013), the optimal model averaging for linear mixed-effects models and generalized linear mixed-effects models (Zhang et al. 2014; Zhang et al. 2015).
In contrast, very little research has been done on the development of model averaging in high-dimensional settings, particularly when the number of predictors is much larger than the sample size.
For example, Ando and Li (2014) developed a novel
model averaging procedure for high-dimensional regression models using a delete-one cross-validation criterion. This effective approach is designed for unsurvival response variables, but not for survival response variables, and the formulation is not applicable
to high-dimensional survival regression models. To the best of our knowledge, there does not seem to exist any established model averaging method for improving forecasting accuracy of survival response variables in high-dimensional sparse regression models.

In this article, we  present a two-step model averaging approach for high-dimensional survival regression.  The first step is to prepare a class of candidate models for averaging. Instead of using the existing variable screening procedures under covariate-independent censoring, we propose a new variable screening procedure through the marginal Buckley-James index to separate the active regressors and the inactive ones and group regressors from the selected predictors
with equal size in each group, which usually implies the
similar size of screening index among the group regressors, to
form regression models with survival response variables. The second step is to find the optimal model weights for averaging a class of pre-constructed survival regression models using a delete-one cross-validation criterion.

The  main contributions of this article are threefold. First, we develop a  variable screening method based on a marginal Buckley-James index (Lai and Ying 1988). This method can easily deal with covariate-dependent censoring  and enjoys  sure screening property under  mild regularity conditions. The superior performance of the new method over the existing variable screening methods is demonstrated by simulations.  Second,  the adapted delete-one cross-validation criterion to survival situations can asymptotically minimize the squared error between the true mean and the predicted value, where the standard  constraint that the weights sum to one is removed. Third, simulation results show that the combination of the new variable screening method for preparing a class of candidate models and the delete-one cross-validation criterion for averaging yields a superior method over the existing methods, including  AIC model averaging, BIC model averaging, Lasso, Group Lasso, SCAD, and MCP.

The remainder of the article is organized as follows. We introduce models and present a two-step predicting procedure in Section \ref{section2}. The new variable screening procedure through a marginal Buckley-James index and its corresponding sure screening property are summarized in Section \ref{section3}.   We introduce the model averaging method and establish the theoretical properties  in Section \ref{section4}.  Simulation  and real data analysis results are reported in Sections \ref{section5} and \ref{section6}, respectively, to assess the performance of the proposed   procedure compared to several existing methods. Some concluding remarks are made in Section \ref{section7}. All technical details are provided in the Appendix.

\section{Models and Methods}\label{section2}
Consider a survival study with $n$ subjects and   $p$ covariates denoted by $X=(x_{1},\cdots,x_{p})^{\T}$, where $p>n$. Let $T$ and $U$ denote the survival and censored times,
and $Y$ and $C$ are the corresponding transformed forms, respectively, such as the popularly-used logarithm, $Y=\log T$, $C=\log U$. The observations consist of $\{Y^*_i, \delta_i, X_i, i=1, \ldots, n\}$, independent copies of $\{Y^*, \delta, X\}$ with $Y^*=Y\land C$ and $\delta=I(Y\le C)$.  The goal is to predict the true mean of survival response variable $Y$ based on the observed data. Suppose that $Y$   takes the following
 semiparametric
linear regression model:
\begin{equation}{\label{DW1}}
Y_i=X_i^{\T}\boldsymbol{\beta}+\epsilon_i, i = 1, \ldots, n,
\end{equation}
where $\boldsymbol{\beta}=(\beta_1,\ldots,\beta_p)^{\T}$ is a vector of unknown regression coefficients and
$\epsilon_i$'s
are independent and identically distributed random errors
with an unknown distribution $F$ such that $E|\epsilon_i|<\infty$ while
$E(\epsilon_i)$ need not to be 0. Further, we
assume that $\epsilon_i$ is independent of $(X_i, C_i)$
and conditional on $X_i$, $Y_i$ and $C_i$ are independent. Let $d$ be the number of nonzero components of the true parameter
vector $\boldsymbol{\beta}_0=
(\beta_{01},\cdots,\beta_{0p})^{\T}.$

For the linear regression model defined in (\ref{DW1}),
we let $\mathcal{D}$= $\{j :\beta_{0j}\neq 0,1\leq j\leq p\}$ be the true sparse model with nonsparsity size $d=|\mathcal{D}|$.
In applications, both $d$ and the set of true predictors $\mathcal{D}$ are unknown.
To improve forecasting accuracy in high-dimensional survival regression, we propose a two-step model averaging procedure as follows.

\noindent {\it Step 1:} Construct candidate models.

We firstly partition the $p$ predictors into $K+1$ groups by the developed new feature screening method which will be presented in Section 3, where the first group has the highest values of the marginal utility and the $(K+1)$th group has values closest to zero. We adopt the strategy of Ando and Li (2014) and remove the $(K+1)$th group, which is expected to contain
only insignificant predictors. Then the first-step model averaging procedure is to construct candidate models.
Denote a set of $K$ candidate models $M_1,\ldots,M_K$ by
 \begin{equation}{\label{DWsub}}
 M_k: \ Y_{i}=\beta_{0k}+\sum\limits_{j\in A_k}x_{ij}\beta_j+\varepsilon_i,\;  i=1, \ldots, n,
 \end{equation}
 where $A_k$ is the index set of predictors in the $k$th candidate model $M_k$ for $k = 1,\ldots,K$, $E\varepsilon_i=0$. Set $p_k=|A_k|+1$, where $|A_k|$ denotes the cardinality of $A_k$.  For estimation of $\beta_{0k}$, $\boldsymbol{\beta}_k=\{\beta_j: j\in A_k \} $, we use the ordinary least squares method.
Here, we assume that $\mathcal{D}\subset \{A_1\cup A_2\cup \cdots\cup A_K\}\subset A$ and $A_k\cap A_j=\phi$ for any $k\neq j$, where $A=\{1,\ldots,p\}$.

\noindent {\it Step 2:} Determine the Optimal Model Weights for Averaging.

To achieve the optimality of model weights, we propose to use a delete-one cross-validation criterion without the classical restriction that  weights sum to one, where the deletion is achieved on subjects and not
on the candidate models. The details are given in Section 4.



\section{Marginal Buckley-James Filter}\label{section3}

When there are many candidate models, it is computationally intensive to evaluate model-averaging estimator
in high-dimensional regression models. It is desirable to adopt a feature screening approach to screen important predictors prior to model averaging in the presence of survival responses.
To the best of our knowledge, all the existing nonparametric and model-based independence screening methods are designed for the case of covariate-independent censoring in the literature. In this section, we provide a new  screening procedure based on model (\ref{DW1}) for a general case where survival variable is allowed to depend on predictors.

\subsection{Screening method}

In this part, we refer to marginal semiparametric linear models as fitting models with componentwise covariates.
Following the analysis of Fan and Song (2010), we conclude that when the design
matrix  is standardized, the ranking by the magnitude of the marginal correlation is in fact the same as the ranking by the magnitude of the marginal Buckley-James estimator. i.e., we assume that the covariates are standardized
to have mean zero and standard deviation one:
$$Ex_j = 0 \;{\rm and}\; Ex_j^2=1, j=1, \ldots, p.$$
Let $$\mathcal{S}(s,t\mid F)=tI(t\leq s)+\frac{\int_s^\infty udF(u)}{1-F(s)}I(t>s),$$ which is a martingale with respect to $\mathcal{G}$, where $\mathcal{G}=\{\mathcal{G}_{t_j}\}$ is a family of increasing $\sigma$-fields and $\mathcal{G}_{t_j}$ is the minimal  $\sigma$-field, whose careful explanations can be seen in Ritov(1990). Let $\epsilon_i(\beta_j^\star)=Y_i- x_{ij}\beta_j^\star$
 and $\zeta_i(\beta_j^\star)=C_i- x_{ij}\beta_j^\star$ be the residuals of survival and censored response in marginal model, where parameter $\beta_j^\star$'s  in the marginal model are introduced to vary from  $\beta_j$'s in the joint working model. By denoting $\upsilon_i(\beta_j^\star)=\min(\zeta_i(\beta_j^\star), \epsilon_i(\beta_j^\star))$, $G_n(\beta_j^\star,u)=\sum_{i=1}^nI(\upsilon_i(\beta_j^\star)\ge u).$
Then, for given $\beta_j^\star$,  we can estimate $F_{\beta_j^\star}$, which is the population distribution of  $\epsilon_i(\beta_j^\star)$ by the
Kaplan-Meier estimator
\begin{equation}\nonumber
\widehat{F}_{\beta_j^\star}(t)=1-\prod_{i:\upsilon_i(\beta_j^\star)\leq t}\left\{1-\frac{1}{G_n(\beta_j^\star,\upsilon_i(\beta_j^\star))}\right\}^{\delta_i},
\end{equation}
and
the marginal Buckley-James estimator
$\widehat{\beta}_j^\star$, for $j=1,\ldots,p$, is defined as the solution of $\Psi_n(\beta_j^\star)=0$, where the
estimating function $\Psi_n(\beta_j^\star)$ is given by
\begin{equation}\label{Dd}
\Psi_n(\beta_j^\star)=n^{-1/2}\sum_{i=1}^nx_{ij}\mathcal{S}(\zeta_i(\beta_j^\star),\epsilon_i(\beta_j^\star)\mid \widehat{F}_{\beta_j^\star}).
\end{equation}

We choose $\hat\beta^\star_j$ as a marginal utility to measure the importance of $X_j$ to survival response variable $Y$, and select a set of variables
$$\widehat{\mathcal{D}}= \{j :|\widehat{\beta}_j^\star|\ge\gamma_n,1\leq j\leq p\},$$
where $\gamma_n$ is a predefined  threshold value. This independence learning under the magnitude of marginal Buckley-James
coefficients can rank the importance of features and dramatically decrease the dimension of the parameter space from $p$ (possibly hundreds of thousands) to a much
smaller number by choosing a large $\gamma_n$, and hence  model averaging in a reduced dimension is much more
feasible.  Next, we discuss the sure screening property for an appropriate $\gamma_n$.


\subsection{Theoretical property}

Define
\begin{eqnarray}\label{indest}
\begin{aligned}
\widetilde{\Psi}_{i}(\beta_j^\star)&=\int I(\zeta_i({\beta_j^\star})\ge u)(x_{ij}-\mathbb{D}_{\beta_j^\star}
(u))W_{F_{\beta_j^\star}}(u)d\mathcal{M}(u,\epsilon_i({\beta_j^\star})\mid F_{\beta_j^\star}),\\ \widetilde{\Psi}_{n}(\beta_j^\star)&=\sqrt{n}\sum_{i=1}^n\widetilde{\Psi}_{i}(\beta_j^\star),
\end{aligned}
\end{eqnarray}
where
\begin{eqnarray*}
\mathcal{M}(s,t\mid F)&=&I(t\leq s)-\frac{\int^s_{-\infty}I(t\ge u)dF(u)}{1-F(u)},\\
W_F(t)&=&t-\frac{\int_t^\infty udF(u)}{1-F(t)},\\
\mathbb{D}_{\beta_j^\star}(u)&=&E(x_j|Y^*-x_j\beta_j^\star\ge u).
\end{eqnarray*}
Correspondingly, we define the population version $\beta_{0j}^\star$ of the
marginal Buckley-James least square estimator: $E\{\widetilde{\Psi}_i(\beta_{0j}^\star)\}=0$, where   $E$ denotes the expectation under the true model.

To  establish the sure independence property of the proposed screening method, we need the following regularity condition.

\begin{itemize}
\item[(C1)] (i). For any $j$, $\sup_{i}|x_{ij}|\leq c$. (ii).  For any $j$, there exists a positive
constant $\eta$ satisfying  $\frac{1}{n}\sum_{i=1}^nE|\widetilde{\Psi}_{i}(\beta_{0j}^\star)|^m\leq \frac{m!}{2}\eta^{m-2},m=2,3,\ldots.$
\end{itemize}

The boundedness in (C1) (i) is a common assumption in regression in high-dimensional settings, and (C1) (ii) is a basic condition for the Bernstein's inequality.

\begin{theorem}
Suppose Condition (C1) holds.  Assume $$\lim_{n\rightarrow\infty}n^{1/2-4\varsigma}\Big\{
\inf_{|\beta_j^\star|\leq\rho,|\beta_j^\star-\beta_{0j}^\star|\ge n^{-\psi}}|\widetilde{\Psi}_n(\beta_j^\star)|\Big\}=\infty$$
for $4\varsigma+\psi>1$ with $\frac{1}{8}\leq\varsigma<1$, and  $\eta t+\sqrt{2t}\gg n^{4\varsigma-1}$ for any fixed value $\eta$ and arbitrary positive $t$, where $\rho$ is a positive constant. Then we have
\begin{itemize}
\item[(i)]
$$P\Big(\max_{1\leq j\leq p}|\widehat{\beta}_j^\star-\beta_{0j}^\star|\ge \eta t+\sqrt{2t}\Big)\leq p\exp(-nt).$$
\item[(ii)] (Sure Screening Property) If additional condition ${\rm cov}(X^{\T}\boldsymbol{\beta}_0,x_j)\ge \eta t+\sqrt{2t}$ for any fixed value $\eta$ and arbitrary positive $t$ and $j=1,\ldots,p,$ holds, then by taking $\gamma_n=c_1(\eta t+\sqrt{2t})$ with $0<c_1\leq 1/2$, we have
    $$P(\mathcal{D}\subset\widehat{\mathcal{D}})\ge 1-d\exp(-\frac{nt}{4}).$$
\item[(iii)] Let $\Sigma=E(XX^{\T})$. Under the same conditions in (ii), $$P\Big(|\widehat{\mathcal{D}}|\leq O\{\lambda_{\max}(\Sigma)/(\eta t+\sqrt{2t})^2\}\Big)\ge 1- d\exp(-\frac{nt}{4}),$$
     where $\lambda_{\max}(\Sigma)$ denotes the maximum eigenvalue of $\Sigma$.
\end{itemize}
\end{theorem}

Since covariates $x_j$'s are bounded, then $\eta$ can be taken as a finite constant. The arbitrary $t$ in the formulation ${\rm cov}(X^{\T}\boldsymbol{\beta}_0,x_j)\ge \eta t+\sqrt{2t}$ is introduced to measure the degree of relation between the linear combination of all covariates and any single covariate, which may degenerate with the dimensionality $p$ or the sample size $n$. Therefore, if $t$  is set as $n^{-2\tau}$ $\tau$ being some positive constant  $0<\tau<1/2$, then Theorem 1 (i) is equivalent to
$$P\Big(\max_{1\leq j\leq p}|\widehat{\beta}_j^\star-\beta_{0j}^\star|\ge \eta n^{-2\tau}+\sqrt{2}n^{-\tau}\Big)\leq p\exp(-n^{1-2\tau}).$$
We find that Buckley-James screening (BJS) method can handle the NP-dimensionality $\log(p)=o(n^{1-2\tau})$ with  $0<\tau<1/2$,
which is consistent with the results of Fan and Lv (2008) in linear regression.
The sure screening property given in Theorem 1 (ii) implies that it is only the size of nonsparse elements $d$ that matters for the purpose
of sure screening, not the dimensionality $p$.
 If we add additional condition $\min_{j\in\mathcal{\mathcal{D}}}|\beta_{0j}^\star|-\max_{j\in\mathcal{\mathcal{D}}^c}|\beta_{0j}^\star|\ge 2(\eta t+\sqrt{2t})$ and other assumptions on sample size and dimensionality, we can conclude ranking consistency
 $$\liminf_{n \rightarrow \infty} ( \min_{j\in \mathcal{D}} |\widehat{\beta}_{j}^\star| - \max_{j \notin \mathcal{D}} |\widehat{\beta}_{j}^\star|  )>0,$$
that demonstrates the values of $|\widehat{\beta}_{j}^\star|$'s corresponding to the active explanatory variables
are always ranked beyond  those corresponding to the inactive ones with a high probability.
Thus,  we can well differentiate the active and inactive explanatory variables by taking an ideal threshold value for model averaging procedure. However, in practice, we adopt the results of Theorem 1 (iii) to specify the number of  informative regressors. If we take $t=n^{-2\tau}$, it follows from Theorem 1 (iii) that
$|\widehat{\mathcal{D}}|\leq O\{n^{2\tau}\lambda_{\max}(\Sigma)\}$ with probability tending to one. Particularly, when $\lambda_{\max}(\Sigma) = O(n^\kappa)$, the size of the selected variables is at the order of $O(n^{2\tau+\kappa})$, which confirms that the number of the selected variables can be controlled effectively. 
Therefore, when
$2\tau+\kappa<1$, the hard thresholding rule considered in Fan and Song (2010)
selecting the top $c\lceil n/\log(n)\rceil$ variables can include the true active predictors with high probability, where $c$ is some constant and $\lceil a \rceil$ denotes the
integer part of real value $a$. We adopt this strategy to select predictors. Then we partition them into $K$ groups to construct $K$ candidate models for model averaging, and the $(K+1)$th group consists of the remaining $p-c\lceil n/\log(n)\rceil$ covariates and is removed in model averaging analysis. The rigorous proof of the determination of the
constant $c$ and $K$ for  the problem is worthy for further investigation.

\section{Optimal Model Averaging Procedure}\label{section4}
In this section, we present an optimal averaging procedure based on the selected $q$ covariates and $K$ candidate models. We still use $X=\{x_j: j\in\widehat{\mathcal{D}}\}$ denoting a vector of $q$-dimensional covariates after the screening procedure is applied, and $\boldsymbol{\beta}=\{\beta_j: j\in\widehat{\mathcal{D}}\}$ being the corresponding regression coefficients.
When the $p$ is not very large, we can simply take $q=p$ and proceed directly with the optimal model averaging procedure.

\subsection{Least squares estimation for model $M_k$}

As $Y_i$ cannot be completely observed due to censoring,
 we impute $Y_i$ under model (\ref{DW1}) but with the selected $q$ covariates  by its conditional expectation given the   observed data as follows:
\begin{eqnarray}\nonumber
Y_i(\boldsymbol{\beta},F)&=&E(Y_i\mid X_i,  Y_i^*, \delta_i)\\\label{imput}
&=&\delta_iY_i^*+(1-\delta_i)\Big\{ X_i^{\T}\boldsymbol{\beta}+\frac{\int_{Y_i^*- X_i^{\T}\boldsymbol{\beta}}^\infty tdF_{\boldsymbol{\beta}}(t)}{1-F_{\boldsymbol{\beta}}(Y_i^*- X_i^{\T}\boldsymbol{\beta})}\Big\}.
\end{eqnarray}
 We adopt the penalized Buckley-James method (Wang et al., 2008) to obtain coefficient  estimate $\widehat{\boldsymbol{\beta}}$ and  the Kaplan-Meier estimator $\widehat{F}=\widehat{F}_{\widehat{\boldsymbol{\beta}}}$.
Then we define $Y_{ni}=Y_i(\widehat{\boldsymbol{\beta}},\widehat{F})$, and get the least square estimator of $p_k$-dimensional vector of unknown parameters  $\boldsymbol{\beta}_k$ in the $k$th candidate model $M_k$:
 \begin{equation}{\label{DWest}}
 \widehat{\boldsymbol{\beta}}_k=(\boldsymbol{X}^{\T}_{k}\boldsymbol{X}_{k})^{-1}\boldsymbol{X}^{\T}_{k}\boldsymbol{Y}_n,
\end{equation}
 where $\boldsymbol{Y}_n=(Y_{n1},\cdots,Y_{nn})^{\T}$, $\boldsymbol{X}_{k}=(\boldsymbol{1},\boldsymbol{x}_{k1},\cdots, \boldsymbol{x}_{kp_k})$, where $\boldsymbol{x}_{kj}$ denotes $j$th column of design matrix in the $k$th candidate model. By using the $k$th candidate model where $k=1,\ldots, K.$, the true conditional mean of $\boldsymbol{Y}=(Y_1,\ldots, Y_n)$ can be estimated by $\widehat{\boldsymbol{\mu}}_{k}=\boldsymbol{X}_k\widehat{\boldsymbol{\beta}}_k$.

\subsection{Optimal  model weight selection for averaging}

After applying the least square method to each candidate model, we achieve a list of predictors $\{\widehat{\boldsymbol{\mu}}_{n1}, \cdots, \widehat{\boldsymbol{\mu}}_{nK}\}$. Now we determine the weight of each candidate model. The $k$th  hat matrix $\boldsymbol{X}_k(\boldsymbol{X}^{\T}_{k}\boldsymbol{X}_{k})^{-1}\boldsymbol{X}^{\T}_k$ of each candidate model is denoted by $H_{k}$.

Let $\boldsymbol{\omega}=(\omega_1, \cdots, \omega_K)$ be the weight vector of the $K$ models and
 $$\mathcal{V}=\{\boldsymbol{\omega}\in[0,1]^K: 0\leq\omega_k\leq1\}.$$
 Here,  the standard restriction $\sum\limits_{k=1}^{K}\omega_k=1$ is removed.  
First, the constructed candidate models proposed in the first step do not seem to be equally competitive. The first few candidate models are likely to be more
informative than the last few candidate models because of the ordering of
regressors by marginal screening utility with the survival response variables. If we fixed the
total weight to shift the weights away from the first few candidate models, it would be non-beneficial for prediction. Second, in the extreme example, the predictors from each candidate model become uncorrelated with each other as well and the optimally combined predictors is the sum of all model predictors when the regressors are uncorrelated with each other and the noise variance is ignorable. Therefore, the optimal weight assignment should be $(1,1,\ldots,)$, which implying that the total weight equals to $K$, not $1$. Third, we do not need the total weight constraint to prove the theoretical results of model averaging procedure.

We denote the true conditional mean of $\boldsymbol{Y}$ on $\boldsymbol{X}$ by $\boldsymbol{\mu}$, where $\boldsymbol{X}=(X_1, \cdots, X_n)^{\T}$.   Then its model average predictor is given by
  \[
\begin{array}{llll}
\widehat{\boldsymbol{\mu}}(\boldsymbol{\omega})&=&\sum\limits_{k=1}^K\omega_k\widehat{\boldsymbol{\mu}}_{k}=\sum\limits_{k=1}^K\omega_k\boldsymbol{X}_k(\boldsymbol{X}^{\T}_{k}\boldsymbol{X}_{k})^{-1}\boldsymbol{X}^{\T}_k\boldsymbol{Y}_n\\ &=&\sum\limits_{k=1}^K\omega_kH_{k}\boldsymbol{Y}_n=H(\boldsymbol{\omega})\boldsymbol{Y}_n,
\end{array}
\]
where $H(\boldsymbol{\omega})=\sum\limits_{k=1}^K\omega_kH_{k}$ is the corresponding weighted hat matrix.

We adopt delete-one cross-validation strategy utilized by
Ando and Li (2014). Let $\tilde{u}_{k}^{(-m)}$ be the predicted value of the $m$th observation from the $k$th model $M_k$, which is derived from the observations except for $(Y_m, \delta_m, {X}_m)$. Define $\tilde{\boldsymbol{\mu}}_{k}=(\tilde{u}_{k}^{(-1)},\ldots,\tilde{u}_{k}^{(-n)})^{\T}$. As shown in Ando and Li (2014), we have $\tilde{\boldsymbol{\mu}}_{k}=\tilde{H}_{k}\boldsymbol{Y}_n$, where $\tilde{H}_{k}$ is the smoothing matrix given by $\tilde{H}_{k}=D_{k}(H_k-I)+I$ and $D_{k}$ is the $n\times n$ diagonal matrix with $j$th diagonal element equal to $(1-h_{kj})^{-1}$, where $h_{kj}$ is the $j$th diagonal element of $H_k$. The delete-one predictor   is given by
$$\tilde{\boldsymbol{\mu}}(\boldsymbol{\omega})=\sum\limits_{k=1}^K\omega_k\tilde{\boldsymbol{\mu}}_{k}=\sum\limits_{k=1}^K\omega_k\tilde{H}_{k}\boldsymbol{Y}_n=\tilde{H}(\boldsymbol{\omega})\boldsymbol{Y}_n,$$
where $\tilde{H}(\boldsymbol{\omega})=\sum\limits_{k=1}^K\omega_k\tilde{H}_{k}$. Thus, the delete-one cross-validation criterion can be written as
  \[
\begin{array}{llll}
\mathcal{M} (\boldsymbol{\omega})&=(\boldsymbol{Y}_n-\tilde{\boldsymbol{\mu}}(\boldsymbol{\omega}))^{\T}(\boldsymbol{Y}_n-\tilde{\boldsymbol{\mu}}(\boldsymbol{\omega}))\\
&=(\boldsymbol{Y}_n-\tilde{H}(\boldsymbol{\omega})\boldsymbol{Y}_n)^{\T}(\boldsymbol{Y}_n-\tilde{H}(\boldsymbol{\omega})\boldsymbol{Y}_n)
.
\end{array}
\]

Then, minimizing the $\mathcal{M}(\boldsymbol{\omega})$ over the space $\mathcal{V}$ yields the selected weights
\begin{equation}{\label{DWCV}}
\begin{aligned}
\widehat{\boldsymbol{\omega}}=\arg\min_{\boldsymbol{\omega}\in\mathcal{V}}\mathcal{M}(\boldsymbol{\omega}).
\end{aligned}
\end{equation}
Then the optimal model averaging  estimator of $\boldsymbol \mu$ is $\widehat{\boldsymbol{\mu}}=\widehat{\boldsymbol{\mu}}(\widehat{\boldsymbol{\omega}})$.

The proposed model averaging estimator $\hat{\boldsymbol\mu}$ of $\boldsymbol\mu$  can be used to predict the restricted mean survival time. For example, suppose the survival time takes  the accelerated failure time (AFT) model $$\log(T)=X^{\T}\boldsymbol{\beta}+\epsilon.$$
Let $U$ be a censoring variable. By setting $Y=\log T$ and $C=\log U$, the observed data consist of $\{Y^*_i=\min(Y_i, C_i), \delta_i=I(Y_i\le C_i), i=1, \ldots, n\}$.
The restricted mean survival time is defined as $u=E\{\min(T,v)\mid X\}$, where $v$ is a pre-specified time point. Write ${\bf u}=(u_1, \ldots, u_n)^{\T}$ with $u_i=E\{\min(T_i,v)|X_i\}$.
Direct calculations yield
$$u_i=\int_0^v S_\epsilon(\log t-\mu_i)dt,$$
where $S_\epsilon(\cdot)$ is the survival function of $\epsilon-E(\epsilon)$, and $\mu_i=E(Y_i|X_i)$. Clearly, one can estimate $S_\epsilon(\cdot)$ by its Kaplan-Meier estimator (KME) $\widehat{S}_\epsilon(\cdot)$ based on the data $\{Y^*_i-\widehat{\mu}_i, \delta_i, i=1,\ldots, n\}.$   Therefore   ${\bf u}=(u_1, \ldots, u_n)^{\T}$ can be estimated by $\hat{\bf u}=(\hat u_1, \ldots, \hat u_n)^{\T}$ with
$$
\hat{u}_i=\int_0^v\widehat{S}_\epsilon(\log t-\widehat{\mu}_i)dt.
$$

\subsection{Theoretical results of model averaging}

 Let $\boldsymbol{\mu}=E(\boldsymbol{Y}|\boldsymbol{X})$.
  We  consider loss function $L(\boldsymbol{\omega})
  =(\boldsymbol{\mu}-\widehat{\boldsymbol{\mu}}(\boldsymbol{\omega}))^{\T}(\boldsymbol{\mu}-\widehat{\boldsymbol{\mu}}(\boldsymbol{\omega}))$. The corresponding risk function is
\begin{equation}{\label{DWrisk}}
R(\boldsymbol{\omega})=E[L(\boldsymbol{\omega})|\boldsymbol{X}].
 \end{equation}

Define $\xi_n=\inf\limits_{\boldsymbol{\omega}\in\mathcal{V}}R(\boldsymbol{\omega})$.
Let $\lambda_{\max}(A)$ be the largest eigenvalue of matrix $A$, and  $\vartheta(A)$ be the maximum diagonal element of matrix $A$. In the following, $B$, $B_k$'s, $c$ and $c_k$'s are finite  constants.   To establish the property of proposed model averaging approach,   we need the following regularity conditions.
\begin{itemize}
\item[(C2)] (i) $\sup\limits_k\frac{\vartheta(H_k)}{p_k}\leq \frac{B_1}{n}$;
(ii) $\sup\limits_k\frac{p_k}{n^{3/4}}\leq B_2$;
(iii) For  some fixed integer $\kappa$, $\frac{K^{4\kappa+2}||\boldsymbol{\mu}||^{2\kappa}}{\xi_n^{2\kappa}}=o_p(1)$;
(iv) $\frac{||\boldsymbol{\mu}||^2}{n}\leq B_3$; (v) $K\sqrt{q
/n}\rightarrow 0.$
   \item[(C3)]  $E[\mathcal{S}(\zeta_i(\boldsymbol{\beta}),\epsilon_i(\boldsymbol{\beta})\mid F_{\boldsymbol{\beta}})-E\{\mathcal{S}(\zeta_i(\boldsymbol{\beta}),\epsilon_i(\boldsymbol{\beta})\mid F_{\boldsymbol{\beta}})\}]^{4\kappa}\leq B_4<\infty.$
\end{itemize}
Condition (C2) is the same as that required in Theorem 2 of Ando and Li (2014).
If the order of $\xi_n$ is $n^{1-\phi}$ with $\phi\ge0$, Condition (C2)(iii)  reduces to $K^{1+0.5\kappa^{-1}}=o_p(n^{(1-2\phi)/4})$.
 Since $\kappa$ is fixed and the term $0.5\kappa^{-1}$ is ignorable, $K$ is allowed to grow to infinity when $\phi<1/2$. 
The sub-Gaussian tail behavior of the error term in Condition (C3)
is an extension of Ando and Li (2014).

\begin{theorem}
Under Conditions (C2)-(C3), we have
\begin{equation}{\label{DWtheorem}}
\frac{L(\widehat{\boldsymbol{\omega}})}{\inf\limits_{\omega\in\mathcal{V}}L(\boldsymbol{\omega})}\rightarrow 1
\end{equation}
in probability.
\end{theorem}

\section{Simulation analysis}\label{section5}
To assess the performance of the proposed two-step model averaging methods, we conducted simulations with two objectives. The first objective is to compare the proposed new screening procedure (BJS) with some  existing ones for constructing candidate models for averaging; these include 
quantile-adaptive screening (QAScen;
He et al. 2013), survival rank correlation screening (RCS$_{\rm cen}$; Song et al. 2014), survival conditional quantile screening (CQS$_{\rm cen}$; Wu and Yin 2015), survival sure independent ranking and screening (SIRS$_{\rm cen}$; Zhou and Zhu 2017).  
The second objective is to compare the proposed model averaging method with the classical weight selection-based model averaging methods and the penalized variable selection methods in the reduced variable space selected by the proposed new screening procedure (BJS); these include model averaging with the Buckley-James based Akaike
information criterion under the restriction
$\sum_{k=1}^{K}\omega_k=1$ (MAIC),  model averaging with the Buckley-James based Bayesian information criterion under the restriction
$\sum_{k=1}^{K}\omega_k=1$ (MBIC),
BJS-based model averaging through the  delete-one cross-validation criterion with the restriction
$\sum_{k=1}^{K}\omega_k=1$ and a pre-specified $K$ (MCV1),  BJS-based model averaging with the  delete-one cross-validation criterion without the restriction
$\sum_{k=1}^{K}\omega_k=1$ and using a pre-specified $K$ (MCV2), BJS-based model averaging with delete-one cross-validation without the restriction $\sum_{k=1}^{K}\omega_k=1$ (MCV3), attempting to optimize the choice of $K$ by the proposed cross-validation;
the Buckley-James based Lasso method (Tibshirani 1996), the Buckley-James based group lasso procedure (Glasso; Yuan
and Lin 2006), the penalized Buckley-James regression by SCAD approach (Fan and Li 2001), and the penalized Buckley-James regression by the MCP approach (Zhang 2010). The selection of tuning parameters in variable selection methods will
influence the performance of estimation. Over a grid of values for the regularization parameter lambda, we specify the optimal tuning parameter for both MCP- or SCAD-penalized regression models in R package $ncvreg$ and  Lasso penalty in R package $glmnet$ by $k$(=10)-fold cross validation. We used the R package
$grplasso$ and used the $k$(=5)-fold cross-validation procedure.

In this simulation study, we suppose that response variable $Y=\log(T)$ and the event time $T$ took the accelerated failure time (AFT) model
$$\log(T_i)=X_i^{\T}\boldsymbol{\beta}+\epsilon_i,$$
where $X_i=(x_{i1},\ldots,x_{ip})$. Here $x_{i1},\ldots,x_{i5}$ were independently generated from Unif(0,1), $(x_{i6},\ldots,x_{ip})$ was generated from a multivariate normal distribution with zero mean and covariance matrix $\Phi=(d_{jl})$ with $d_{jl}=0.7^{|j-l|}$, and censoring variable $C_i$ was generated from $C_i=X_i^{\T}\boldsymbol{\theta}+\varepsilon_i$,
 $\epsilon_i$ and $\varepsilon_i$ were assumed to follow the normal distribution $\mathcal{N}(0,0.8^2)$.   Set sample size $n=200$, number of regressors
$p=2000$, the true coefficients $\boldsymbol{\beta}=(3,1.5,0,0,2,\boldsymbol{0}_{p-5})$, and $\boldsymbol{\theta}=(0,0,-4,-4,\boldsymbol{0}_{p-4})$. We consider two censoring proportions: 45$\%$ and 65$\%$.

To prepare candidate models for averaging, we ordered the regressors for grouping using the proposed BJS and the above mentioned screening methods.  
We separated the first $\lceil n/\log(n)\rceil\approx36$ active predictors by each screening method, and set $K=6$ to yield a class of 6 candidate models, each with 6 regressors for MAIC, MBIC, MCV1 and MCV2. MCV3 method checks the performance of the BJS-based model averaging method depend on the optimal $K$ from the considered number $K = 1, 2, 3, 6, 9, 18$.
Then, we conducted the proposed model averaging procedure for the  candidate models from each screening approach.
We obtained the penalized estimates using the Buckley-James-type least square objective function with different penalty functions (Lasso, SCAD, MCP, Glasso). For Glasso, we regard the covariates in each candidate model as a group.

To measure the performance of all screening methods, we considered the minimum model size $\mathbb{M}$, that is, the minimum number of predictors needed to include all the active predictors.
Table 1 presents the median and the standard  error of $\mathbb{M}$ over 200 replications. BJS shows the best performance in terms of median equalling 3 and the smallest standard deviation. Other screening methods actually cannot  work under covariate-dependent censoring. We used  MSE $\frac{1}{n}(\boldsymbol{\mu}-\widehat{\boldsymbol{\mu}})^{\T}
(\boldsymbol{\mu}-\widehat{\boldsymbol{\mu}})$ as the predicting performance measure.
\begin{table}[t!] 
\caption{Simulation results of $\mathbb{M}$ by different screening methods. The numbers in the table are medians and standard  deviation (SE) of 200 replicates.}
\label{tab:simulation}\par
\vskip .2cm
\centerline{\tabcolsep=3truept\begin{tabular}{|rcrrrrrrrrr|} \hline
&& \multicolumn{2}{c}{CR$=45\%$}&&\multicolumn{2}{c}{CR$=65\%$}&&&&
\\[3pt] \hline
Method  && Median  & SD&& Median  & SD &&&&
\\ \hline
BJS && 3 & 0.2&& 5 & 0&&&&\\
RCS$_{\rm cen}$  &&  6 & 41.5&& 11 & 145.0&&&&\\
CQS(0.5)$_{\rm cen}$  && 8 & 122.9&& 14 & 174.5&&&&\\
SIRScen && 5 &2.89&& 11 & 8.96&&&&\\
QAScen(0.5) &&25&32.4 &&78  &136.3&&&&
\\ \hline
\end{tabular}}\vskip .2cm
\noindent{\it Note}: CR denotes the censoring rate; 0.5 means the quantile used in the corresponding screening procedure.
\end{table}


\begin{figure}
  \centering
 \includegraphics[width=15cm,height=7cm]{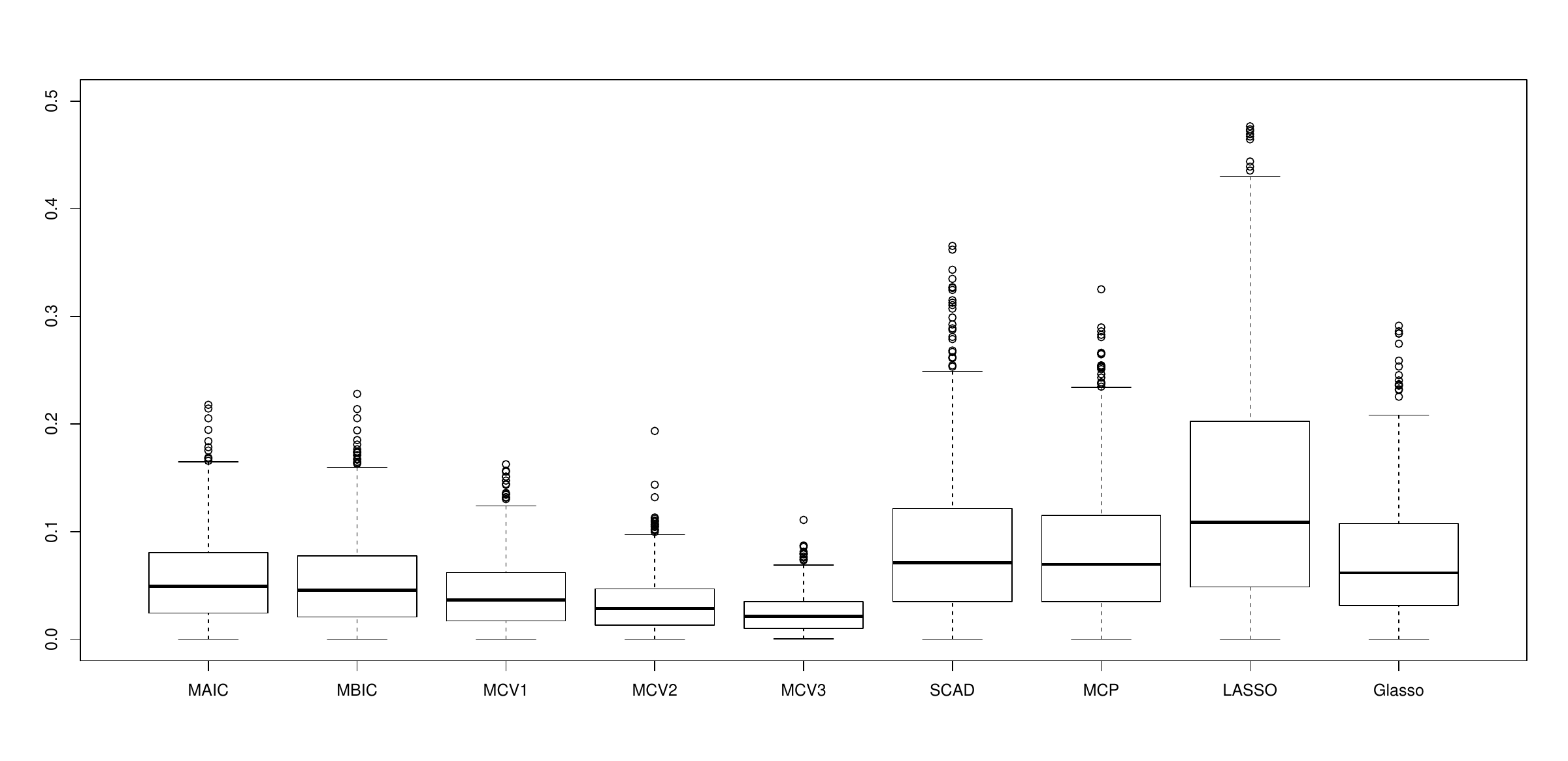}\\
  \includegraphics[width=15cm,height=7cm]{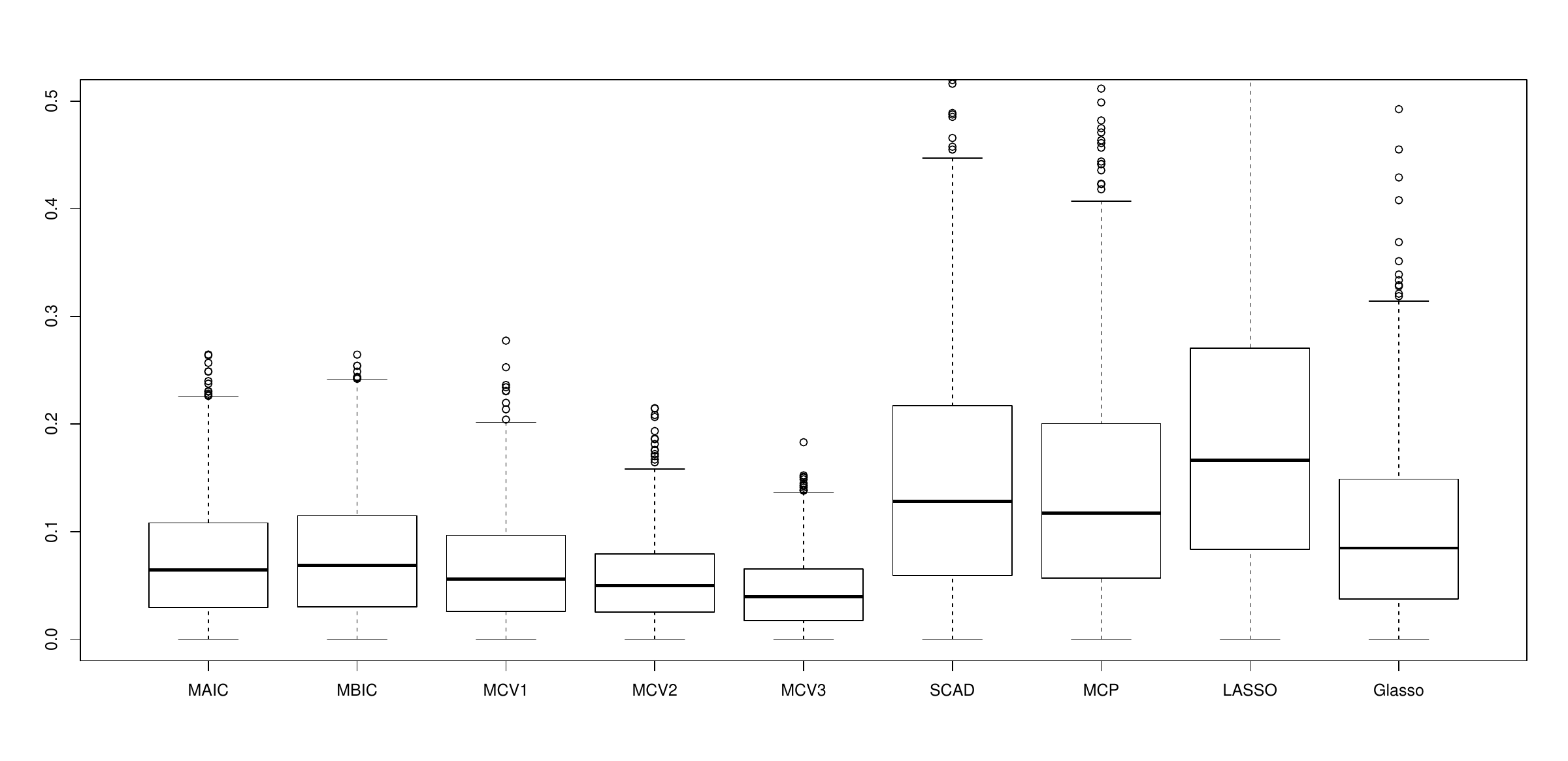}\\
  \caption{ Comparison of the proposed model averaging method with other  model averaging methods and the penalized methods based on simulated data sets under CR=45\%(upper figure) and CR=65\%(lower figure).
  }\label{tex}
\end{figure}


Figure 1 display the boxplots of MSE values for our proposed model averaging procedure and other methods in the reduced dimension through BJS screening procedure based on the same 200 simulated data sets under two censoring ratios. The simulation results demonstrate that our proposed model averaging procedure has the superior  performance over the penalized methods and other model averaging procedures. The proposed model averaging procedure MCV3, selecting the optimal $K$ based on the developed cross-validation shows the best prediction in terms of the smallest median and standard deviation. Meanwhile, high censoring rate wholly destroys the prediction performance.


\section{Empirical study}\label{section6}

In this section, we applied the proposed method to   the  mantle cell lymphoma (MCL) microarray data,  which is available from the web site http://
llmpp.nih.gov/MCL.
The primary goal of this study was to identify genes that have significant
influence on patients' survival or health risk and assess the prediction performance of the proposed model averaging procedure. Among $101$
untreated patients with no history of previous lymphoma, $92$ were classified
as having MCL based on the morphologic and immunophenotypic criteria. The data  includes   the
expression values of 6312 genes  for each patient.
During the follow-up, $64$ patients died of MCL and the other 28 patients
were survival. Here, the sample size is $n=92$, the number of predictors is $p = 6312$ and the censoring rate is $36\%$.

Suppose that the survival time $T$ follows the AFT model considered in our simulations. We applied the proposed BJS independence screening procedure to the survival MCL data and took the first $3\lceil\frac{n}{\log(n)}\rceil=60$ ranked genes as active predictors. Then, we set $K=6$ to yield a class of 6 candidate models, each with 10 genes for MAIC, MBIC, MCV1 and MCV2. MCV3 selects the optimal $K$ from the considered number $K = 1, 2, 3, 6, 10, 15, 20$ to conduct model averaging prediction.


To evaluate the forecasting performance of various methods, we adopt bootstrap strategy  and let $A=\{i: \mbox{observation} \; i \; \mbox{is resampled}\}$ as the index set of observations resampled and ${N}_i$ denotes the resampled number of $i$. We used the averaged squared forecasting errors (ASPE)
$${\rm ASPE}=\frac{1}{\sum_{i \in A }{N}_i\delta_i}\sum_{i \in A }{N}_i\delta_i(Y_{i}-\widehat\mu_i)^2$$
as the forecasting performance measure for each method. We calculated the ASPE from the estimation results by applying the proposed optimal weight selection method to the obtained candidate models based on the randomly selected subset. We also calculated the ASPEs by using MCV, 
SCAD, MCP and LASSO as described in Section 5. Figure 2 shows the boxplots of ASPE values obtained from these methods based on 200 replications. The forecasting results demonstrate the superior performance of the proposed model averaging method  over the classical weight selection methods and the penalized variable selection methods for the MCL data.

\begin{figure}
  \centering
  \includegraphics[width=15cm,height=7cm]{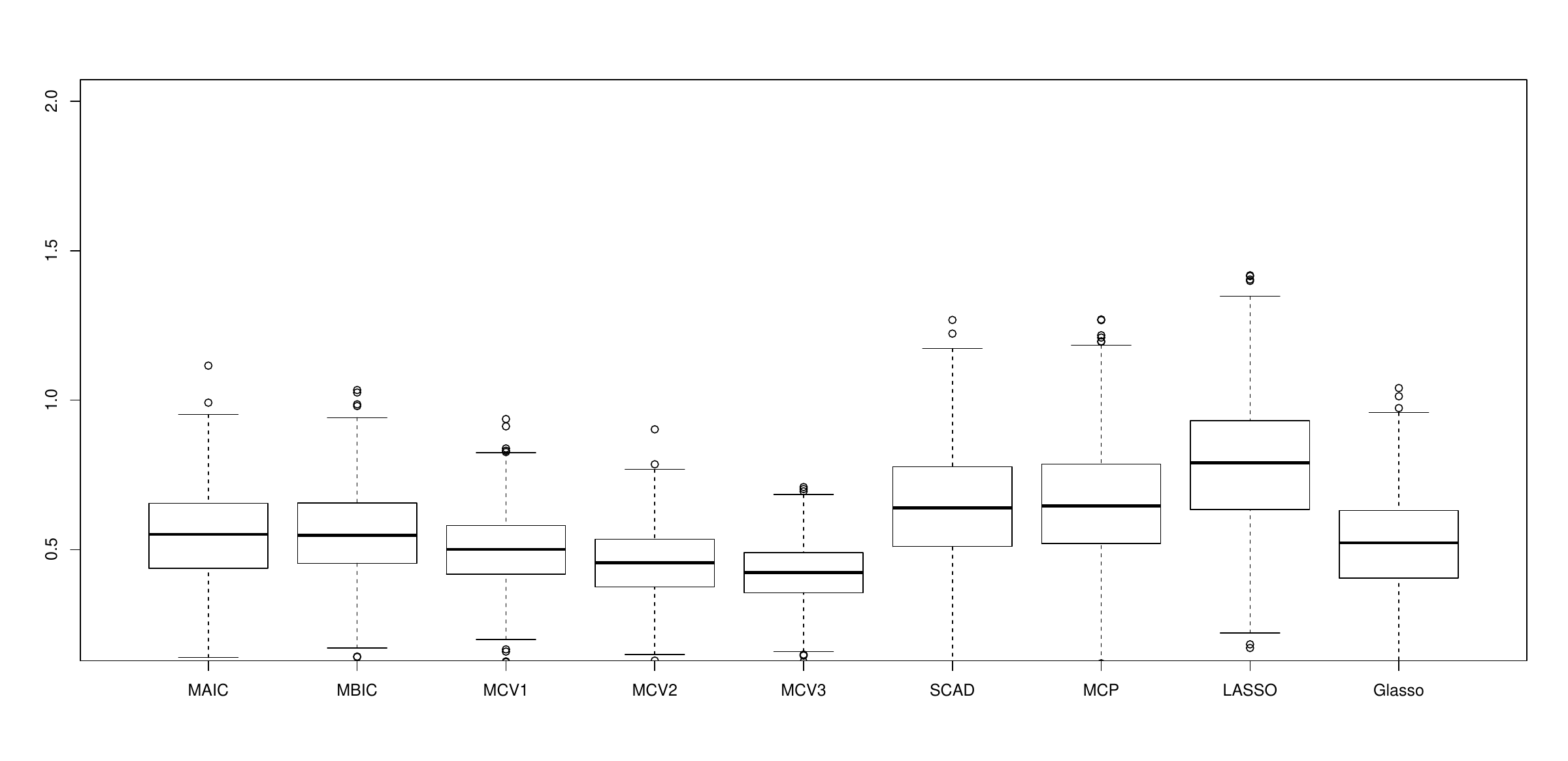}\\
  \caption{ Boxplots for the forecasting performance of the proposed model averaging method and the penalized methods based on the MCL data.
  }\label{tex}
\end{figure}

\section{Conclusions}\label{section7}

For the analysis of high-dimensional data, many penalized methods such as LASSO, SCAD and MCP, have been developed. These penalized procedures, however, may not perform well  for a very large number of covariates because ultrahigh dimensionality brings
  simultaneous challenges of computational expediency,
statistical accuracy and algorithmic stability (Fan et al. 2009). In this article, to address the new challenges from ultrahigh-dimensional data in the presence of censoring, we  have proposed a two-step model averaging  procedure for improving forecasting accuracy. The first step is to construct candidate models for averaging, while the second step is to find the optimal model weights for predicting. For the first step, to accommodate censoring
 in the ultrahigh-dimensional survival data, we have designed a new  screening procedure based on a marginal Buckley-James filter and established its  sure screening property under very weak regularity conditions. The proposed screening approach still work under covariate-dependent censoring.
For the second step, we give an estimation of the survival response then an ordinary least squares method  can be used to estimate regression parameters for each candidate model subject to  censoring, and a delete-one cross-validation criterion has been adapted to determine the optimal model weights, where the standard constraint that weights sum to one has been successfully removed.  The superior performance of the proposed model averaging approach over the classical model selection methods and the penalized methods has been demonstrated through simulation studies and real data analysis.
Further research is to extend the proposed model averaging method to other high-dimensional
survival models such as the Cox model and the additive hazards model.

\section*{Acknowledgments}
The authors are grateful to the Editor, an Associate Editor and two referees for their valuable suggestions and
comments that greatly improved the manuscript. The research was supported by the National Natural Science Foundation of China
[grant number 11901352], the Natural Science Foundation of Shandong Province [grant number
ZR2019BA017], Social Science Foundation of Shandong Province [grant number 19DTJJ03] and the
Young Scholars Program of Shandong University [YSPSDU: 11020088964008].


\newpage
\section*{References}

\begin{description}

\item Ando, T., and Li, K. C. (2014), ``A Model-Averaging Approach for High-dimensional Regression,'' {\sl Journal of the American Statistical Association}, 109, 254-265.

\item Ando, T., and Li, K. C. (2017), ``A Weight-relaxed Model Averaging Approach For
High-dimensional Generalized Linear Models,'' {\sl The Annals of Statistics}, 45, 2654-2679.

\item Ando, T., and Tsay, R. (2010), ``Predictive Likelihood for the Bayesian
Model Selection and Averaging,'' {\sl  International Journal of Forecasting}, 26, 744-763

\item Antoniadis, A., Fryzlewicz, P., and Letu$\acute{e}$, F. (2010), ``The Dantzig Selector in Cox's Proportional Hazards Model,'' {\sl Scandinavian Journal of Statistics}, 37, 531-552.

\item Bradic, J., Fan, J., and Wang, W. (2011), ``Penalized Composite Quasi-likelihood for Ultrahigh-Dimensional Variable Selection,'' {\sl Journal of the Royal Statistical Society} Series B, 73, 325-349.

\item Burnham, K., and Anderson, D. (2002), ``Model selection and multimodel inference: a practical information-theoretic approach,'' 2nd edn. Springer, New York.

\item Candes, E., and Tao, T. (2007), ``The Dantzig Selector: Statistical Estimation When p is Much Larger than n,'' {\sl The Annals of Statistics}, 2313-2351.

\item Cui, H., Li, R., and Zhong, W. (2015), ``Model-Free Feature Screening for Ultrahigh Dimensional Discriminant Analysis,'' {\sl Journal of the American Statistical Association},
    110, 630-641.

\item Ding, Y., and Nan, B. (2011), ``A Sieve M-theorem for Bundled Parameters in Semiparametric Models, with Application to the Efficient Estimation in a Linear Model for Censored Data,'' {\sl The Annals of Statistics}, 39, 2795.

\item Dirick, L., Claeskens, G., and Baesens, B. (2017). ``Time to Default in Credit Scoring Using Survival Analysis: A Benchmark Study,'' {\sl Journal of the Operational Research Society}, 68, 652-665.

\item Fan, J., Feng, Y., and Song, R. (2011), ``Nonparametric Independence Screening in Sparse Ultra-High Dimensional Additive Models,'' {\sl Journal of the American Statistical Association}, 109, 1270-1284.

\item Fan, J., and Li, R. (2001), ``Variable Selection via Nonconcave Penalized Likelihood and Its Oracle Properties. {\sl Journal of the American Statistical Association}, 96, 1348-1360.

\item Fan, J., and Li, R. (2002), ``Variable Selection for Cox's Proportional Hazards Model and Frailty Model,'' {\sl Annals of Statistics}, 30, 74-99.

\item Fan, J., and Lv, J. (2008), ``Sure Independence Screening for Ultrahigh Dimensional Feature Space,'' {\sl Journal of the Royal Statistical Society}, Series B, 70, 849-911.

\item Fan, J., Samworth, R., and Wu, Y. (2009), ``Ultrahigh dimensional feature selection: beyond the linear model,''  {\sl Journal of Machine Learning Research}, 10, 2013-2038.

\item Hansen, B. E. (2007), ``Least Squares Model Averaging,'' {\sl Econometrica}, 75, 1175-1189.

\item Hansen, B. E. and Racine, J. (2012), ``Jackknife Model Averaging,'' {\sl Journal of Econometrics}, 167, 38-46.

\item He, X., Wang, L., and Hong, H. G. (2013), ``Quantile-Adaptive Model-Free Variable Screening for High-Dimensional Heterogeneous Data,'' {\sl Annals of Statistics}, 41, 342-369.

\item Hjort, N. L., and Claeskens, G. (2003), ``Frequentist Model Average Estimators,'' {\sl Journal of the American Statistical Association}, 98, 879-899.




\item Lai, T. L., and Ying, Z. (1988), ``Stochastic integrals of empirical-type processes with applications to censored regression,'' {\sl Journal of Multivariate Analysis}, 27, 334-358.

\item Huang, J., Ma, S., and Xie, H. (2006), ``Regularized Estimation in the Accelerated Failure Time Model with High-Dimensional Covariates,'' {\sl Biometrics}, 62, 813-820.

\item Li, K. C. (1986), ``Asymptotic Optimality of CL and Generalized Cross-Validation in Ridge Regression With Application to Spline Smoothing,''
{\sl The Annals of Statistics}, 14, 1011-1112.



\item Leng, C., and Ma, S. (2007), ``Path Consistent Model Selection in Additive Risk Model via Lasso,'' {\sl Statistics in medicine}, 26, 3753-3770.



\item Liang, H., Zou, G., Wan, A. T. K., and Zhang, X. (2011), ``Optimal Weight Choice for Frequentist Model Average Estimators,'' {\sl Journal of the American Statistical Association}, 106, 1053-1066.

\item Lin, W., and Lv, J. (2013), ``High-Dimensional Sparse Additive Hazards Regression,'' {\sl Journal of the American Statistical Association}, 108, 247-264.


\item Liu, Q. and Okui, R. (2013), ``Heteroskedasticity-Robust $C_p$ Model Averaging,'' {\sl The Econometrics Journal}, 16, 463-472.

\item Mai, Q., and Zou, H. (2015), ``The Fused Kolmogorov Filter: A Nonparametric Model-Free Screening Method,'' {\sl The Annals of Statistics}, 43, 1471-1497.

\item Martinussen, T., and Scheike, T. H. (2009), ``Covariate Selection for the Semiparametric Additive Risk Model,'' {\sl Scandinavian Journal of Statistics}, 4, 602-619.

\item Newbold, P., and Granger, C. W. J. (1974), ``Experience With Forecasting
Univariate Time Series and the Combination of Forecasts'' (with discussion), {\sl Journal of the Royal Statistical Society}, Series A, 137, 131-149.

\item Pablo Montero-Manso, George Athanasopoulos, Rob J. Hyndman,
Thiyanga S. Talagala. (2020), ``FFORMA: Feature-based forecast model averaging'',
{\sl International Journal of Forecasting}, 36, 86-92.

\item Ritov, Y. (1990), ``Estimation in a linear regression model with censored data'',
{\sl The Annals of Statistics}, 18, 303-328.

\item Rosenwald, A., Wright, G., Wiestner, A., Chan, W. C., Connors, J. M., Campo, E., and Chiorazzi, M. (2003), ``The Proliferation Gene Expression Signature is a Quantitative Integrator of Oncogenic Events that Predicts Survival in Mantle Cell Lymphoma,'' {\sl Cancer cell}, 3, 185-197.

\item Stepanova, M., and Thomas, L. C. (2002), ``Survival Analysis Methods for Personal Loan Data,'' {\sl Operations Research}, 50, 277-289.

\item Stute W. (1996), ``Distributional Convergence under Random Censorship When Covariables are Present,'' {\sl Scandinavian Journal of Statistics}, 461-471.

\item Tang, N., Yan, X., and Zhao, X. (2020), ``Penalized generalized empirical likelihood with a diverging number of general estimating equations for censored data,'' {\sl Annals of Statistics}, 48, 607-627.

\item Tibshirani, R. (1996), ``Regression Shrinkage and Selection Via the
Lasso,'' {\sl Journal of the Royal Statistical Society}, Series B, 58, 267-288.

\item Tibshirani, R. (1997), ``The Lasso Method for Variable Selection in the Cox Model,'' {\sl Statistics in medicine}, 16, 385-395.

\item Wan, A. T. K., Zhang, X., and Zou, G. (2010), ``Least Squares Model
Averaging by Mallows Criterion,'' {\sl Journal of Econometrics}, 156, 277-283.

\item Wang, S., Nan, B., Zhu, J.,  and Beer, D. G. (2008), ``Doubly Penalized Buckley-James Method for Survival Data with High-Dimensional Covariates,''
     {\sl Biometrics}, 64, 132-140.

\item Whittle, P. (1960), ``Bounds for the Moments of Linear and Quadratic Forms
in Independent Variables,'' {\sl Theory of Probability and its Applications}, 5,
302-305

\item Xie, J., Yan X., and Tang, N. (2020), ``A model-averaging method for high-dimensional regression with missing responses at random,'' {\sl Statistica Sinica},  DOI:10.5705/ss.202018.0297

\item Yan X., Yin G., and Zhao X. (2020), ``Subgroup analysis in censored linear regression,'' {\sl Statistica Sinica}, DOI:10.5705/ss.202018.0319

\item Yuan, M., and Lin, Y. (2006), ``Model Selection and Estimation in Regression
With Grouped Variables,'' {\sl Journal of the Royal Statistical Society}, Series B, 68, 49-67.

\item Yuan, Z. and Yang, Y. (2005), ``Combining Linear Regression Models: When and How?'' {\sl Journal of the American Statistical Association}, 100, 1202-1214.

\item Zhang, C.-H. (2010), ``Nearly Unbiased Variable Selection Under Minimax
Concave Penalty,'' {\sl The Annals of Statistics}, 38, 894-942.

\item Zhang, H.-H., and Lu, W. (2007), ``Adaptive Lasso for Cox's Proportional Hazards Model,'' {\sl Biometrika}, 94, 691-703.

\item Zhang, X., Zou, G., and Liang, H. (2014), ``Model Averaging and Weight Choice in Linear Mixed-Effects Models,'' {\sl Biometrika}, 101, 205-218.

\item Zhang, X., Yu, D., Zou, G., and Liang, H. (2016), ``Optimal Model Averaging Estimation for Generalized Linear Models and Generalized Linear Mixed-Effects Models,'' {\sl Journal of the American Statistical Association}, 111, 1775-1790.

\item Zhou, M. (1992), ``M-estimation in Censored Linear Models,'' {\sl Biometrika}, 79, 837-841.

\item  Zhu, L. P., Li, L., Li, R., and Zhu, L. X. (2011). ``Model-Free Feature Screening for Ultrahigh-Dimensional Data,''  {\sl Journal of the American Statistical Association}, 106, 1464-1475.


\item Zou, H. (2006). ``The Adaptive Lasso and Its Oracle Properties,'' {\sl Journal of the American Statistical Association}, 101, 1418-1429.

\item Zou, H. (2008), ``A Note on Path-Based Variable Selection in the Penalized Proportional Hazards Model,'' {\sl Biometrika}, 95, 241-247.

\end{description}
\newpage
\appendix
\section*{ Appendix }
\vspace{.1in}
\renewcommand{\theequation}{A.\arabic{equation}}
\setcounter{equation}{0}
\noindent
The marginal Buckley-James type least squares estimating function for the  estimator $\widehat{\beta}_j^\star$ in (\ref{Dd}) is
\begin{equation}\label{Psin}
\Psi_n(\beta_j^\star)=n^{-1/2}\sum_{i=1}^nx_{ij}\mathcal{S}(\zeta_i({\beta_j^\star}),\epsilon_i({\beta_j^\star})\mid \widehat{F}_{\beta_j^\star}),
\end{equation}
where $\mathcal{S}(s,t\mid F)=tI(t\leq s)+\frac{\int_s^\infty udF(u)}{1-F(s)}I(t>s)$.
Define
\begin{equation}
\widetilde{\Psi}_n(\beta_j^\star)=n^{-1/2}\sum_{i=1}^n\int I(\zeta_i({\beta_j^\star})\ge u)(x_{ij}-\mathbb{D}_{\beta_j^\star}
(u))W_{F_{\beta_j^\star}}(u)d\mathcal{M}(u,\epsilon_i({\beta_j^\star})\mid F_{\beta_j^\star}),
\end{equation}
where
\begin{eqnarray*}
\mathcal{M}(s,t\mid F)&=&I(t\leq s)-\frac{\int^s_{-\infty}I(t\ge u)dF(u)}{1-F(u)},\\
W_F(t)&=&t-\frac{\int_t^\infty udF(u)}{1-F(t)},\\
\mathbb{D}_{\beta_j^\star}(u)&=&E(x_j|Y^*-x_j\beta_j^\star\ge u).
\end{eqnarray*}

\begin{lemma}\label{lam0} For a given small constant $\varepsilon$,
\begin{itemize}
\item[(i)] $\sup\{\big|W_{\widehat{F}_{\beta_j^\star}}(t)
    -W_{F_{\beta_j^\star}}(t)\big|: |\beta_j^\star|\leq \rho,\sum_{i=1}^nI(\upsilon_i({\beta_j^\star})\ge s)\ge \frac{cn^{1-\varsigma}}{2}, t\leq s\leq \infty\}=O(n^{-1/2+4\varsigma+\varepsilon})$ a.s..

\item[(ii)] $\sup\{|n^{-1}\sum_{i=1}^n[\delta_i
x_{ij}-\delta_i\mathbb{D}_{\beta_j^\star}(\epsilon_i({\beta_j^\star}))]|: |\beta_j^\star|\leq\rho\}  
=O(n^{-1/2+\varepsilon})$ a.s..

\item[(iii)] $\sup\{|
\widehat{\mathbb{D}}_{\beta_j^\star}(u)-\mathbb{D}_{\beta_j^\star}(u)|: u\leq \infty, |\beta_j^\star|\leq\rho\}=O(n^{-1/2+\varepsilon})$ a.s., where $\widehat{\mathbb{D}}_{\beta_j^\star}(u)={\sum_{i=1}^nx_{ij}I(\upsilon_i({\beta_j^\star})\ge u)}/{\sum_{i=1}^nI(\upsilon_i({\beta_j^\star})\ge u)}$.
\end{itemize}
\end{lemma}

\vspace{.1in}
\noindent {\large\bf Proof of Lemma \ref{lam0}}\\
\vspace{.1in}
By Lemma 2 of Lai and Ying (1991), we have
$$\sup\Big\{\big| \frac{\int_t^{\infty}sd\widehat{F}_{\beta_j^\star}(s)}{1
-\widehat{F}_{\beta_j^\star}(t)}-\frac{\int_t^{\infty}sdF_{\beta_j^\star}(s)}{1
-F_{\beta_j^\star}(t)}\big|:|\beta_j^\star|\leq \rho,t\leq s\leq \infty,
\sum_{i=1}^nI(\upsilon_i(\beta_j^\star)\ge s)\ge \frac{cn^{1-\varsigma}}{2}\Big\}
=O(n^{-1/2+4\varsigma+\varepsilon})$$ a.s. for every
$0\leq\varsigma<1$ and $\varepsilon>0$, and thus Lemma \ref{lam0} (i) holds.
We obtain Lemma 1 (ii) using $\mathbb{D}_{\beta_j^\star}(u)=E(x_j\mid Y^*-x_j\beta_j^\star\ge u)=E(x_j\mid Y-x_j\beta_j^\star=u,\delta=1)$
(Lai and Ying 1991) and
$$E(\delta_iX_{ij})=
E[\delta_i\mathbb{D}_{\beta_j^\star}(\epsilon_i(\beta_j^\star))].$$
We conclude Lemma \ref{lam0} (iii) from the definition of $\mathbb{D}_{\beta_j^\star}(u)=E(x_j\mid Y^*-x_j\beta_j^\star\ge u).$

\begin{lemma}\label{lam1}
$\sup_{|\beta_j^\star|\leq\rho}|\Psi_n(\beta_j^\star)
-\widetilde{\Psi}_n(\beta_j^\star)|=O(n^{-1/2+3\varsigma+2\varepsilon})$ a.s.
\end{lemma}

\vspace{.1in}
\noindent {\large\bf Proof of Lemma \ref{lam1}}\\
\vspace{.1in}
By Proposition 3.2 of Ritov (1990), (\ref{Psin}) is equivalent to
$$\Psi_n(\beta_j^\star)=n^{-1/2}\sum_{i=1}^n\int I(\zeta_i(\beta_j^\star)\ge u)(x_{ij}-\widehat{\mathbb{D}}_{\beta_j^\star}
(u))W_{\widehat{F}_{\beta_j^\star}}(u)d\mathcal{M}(u,\epsilon_i(\beta_j^\star)\mid \widehat{F}_{\beta_j^\star}).
$$
Therefore, $\Psi_n(\beta_j^\star)-\widetilde{\Psi}_n(\beta_j^\star)
=J_{1n}(\beta_j^\star)+J_{2n}(\beta_j^\star)+J_{3n}(\beta_j^\star)$, where
\begin{eqnarray*}
&&J_{1n}(\beta_j^\star)=n^{-1/2}\sum_{i=1}^n\int I(\zeta_i(\beta_j^\star)\ge u)x_{ij}\{W_{\widehat{F}_{\beta_j^\star}}(u)-W_{F_{\beta_j^\star}}(u)\}d\mathcal{M}(u,\epsilon_i(\beta_j^\star)\mid \widehat{F}_{\beta_j^\star}),\\
&&J_{2n}(\beta_j^\star)=n^{-1/2}\sum_{i=1}^n\int I(\zeta_i(\beta_j^\star)\ge u)(x_{ij}-\widehat{\mathbb{D}}_{\beta_j^\star}
(u))W_{F_{\beta_j^\star}}(u)d\{\mathcal{M}(u,\epsilon_i(\beta_j^\star)\mid \widehat{F}_{\beta_j^\star})-\mathcal{M}(u,\epsilon_i(\beta_j^\star)\mid F_{\beta_j^\star})\},\\
&&J_{3n}(\beta_j^\star)=n^{-1/2}\sum_{i=1}^n\int I(\zeta_i(\beta_j^\star)\ge u)\{\mathbb{D}_{\beta_j^\star}(u)-\widehat{\mathbb{D}}_{\beta_j^\star}
(u)\}W_{F_{\beta_j^\star}}(u)d\mathcal{M}(u,\epsilon_i(\beta_j^\star)\mid F_{\beta_j^\star}).
\end{eqnarray*}
For $J_{1n}$, we consider the process
$$L_n(\beta_j^\star)=J_{1n}(\beta_j^\star)-Q_{1n}(\beta_j^\star),$$
where $Q_{1n}(\beta_j^\star)=n^{-1/2}\sum_{i=1}^n\int I(\zeta_i(\beta_j^\star)\ge u)\{W_{\widehat{F}_{\beta_j^\star}}(u)-W_{F_{\beta_j^\star}}(u)\}
\{\mathbb{D}_{\beta_j^\star}(u)-\widehat{\mathbb{D}}_{\beta_j^\star}(u)\}
dN_i(\beta_j^\star,u),$ and
$N_i(\beta_j^\star,u)=I(\epsilon_i(\beta_j^\star)\leq u).$
By Lemma \ref{lam0} (i) and (ii), we have
\begin{eqnarray*}
\big|L_n(\beta_j^\star)\big||&=&\big|n^{-1/2}\sum_{i=1}^n\{W_{\widehat{F}_{\beta_j^\star}}(\epsilon_i(\beta_j^\star))-W_{F_{\beta_j^\star}}(\epsilon_i(\beta_j^\star))\}\\
&&\times\big[x_{ij}-\{\mathbb{D}_{\beta_j^\star}(\epsilon_i(\beta_j^\star))-\widehat{\mathbb{D}}_{\beta_j^\star}(\epsilon_i(\beta_j^\star))\}-\sum_{k=1}^nX_{kj}I(\upsilon_k(\beta_j^\star)\ge\epsilon_i(\beta_j^\star))\frac{d\widehat{F}_{\beta_j^\star}(\epsilon_i(\beta_j^\star))}{1-\widehat{F}_{\beta_j^\star}(\epsilon_i(\beta_j^\star))}\big]\delta_i\big|\\
&=&\big|n^{-1/2}\sum_{i=1}^n\{W_{\widehat{F}_{\beta_j^\star}}(\epsilon_i(\beta_j^\star))-W_{F_{\beta_j^\star}}(\epsilon_i(\beta_j^\star))\}\big[
x_{ij}-\mathbb{D}_{\beta_j^\star}(\epsilon_i(\beta_j^\star))\big]\delta_i\big|\\
&\leq&\sup_{|\beta_j^\star|\leq\rho,t\leq \infty}\big|W_{\widehat{F}_{\beta_j^\star}}(t)-W_{F_{\beta_j^\star}}(t)\big|\big|n^{-1/2}\sum_{i=1}^n\{
\delta_iX_{ij}-\delta_i\mathbb{D}_{\beta_j^\star}(\epsilon_i(\beta_j^\star))\}\big|\\
&=&O(n^{-1/2+3\varsigma+2\varepsilon})\;a.s.
\end{eqnarray*}
On the other hand, using Lemma  \ref{lam0} (i) and (iii), we have
\begin{eqnarray*}
|Q_{1n}(\beta_j^\star)|&\leq& |n^{-1/2}\sum_{i=1}^n\int \{W_{\widehat{F}_{\beta_j^\star}}(u)-W_{F_{\beta_j^\star}}(u)\}\{\mathbb{D}_{\beta_j^\star}(u)-\widehat{\mathbb{D}}_{\beta_j^\star}(u)\}dN_i(\beta_j^\star,u)|\\
&=&|n^{-1/2}\sum_{i=1}^n \delta_i\{W_{\widehat{F}_{\beta_j^\star}}(\epsilon_i(\beta_j^\star))-W_{F_{\beta_j^\star}}(\epsilon_i(\beta_j^\star))\}\{\mathbb{D}_{\beta_j^\star}(\epsilon_i(\beta_j^\star))-\widehat{\mathbb{D}}_{\beta_j^\star}(\epsilon_i(\beta_j^\star))\}|\\
&\leq&
\sup_{|\beta_j^\star|\leq\rho,t\leq \infty}\big|W_{\widehat{F}_{\beta_j^\star}}(t)-W_{F_{\beta_j^\star}}(t)\big||n^{-1/2}\sum_{i=1}^n (\mathbb{D}_{\beta_j^\star}(\epsilon_i(\beta_j^\star))-\widehat{\mathbb{D}}_{\beta_j^\star}(\epsilon_i(\beta_j^\star)))|\\
&=&O(n^{-1/2+3\varsigma+2\varepsilon})\; a.s..
\end{eqnarray*}
Therefore, $$|J_{1n}(\beta_j^\star)|\leq |L_{n}(\beta_j^\star)|+|Q_{1n}(\beta_j^\star)|=O(n^{-1/2+3\varsigma+2\varepsilon})\;a.s..$$
For $J_{2n}$, by $W_{F_{\beta_j^\star}}(u)\leq \infty$ and Lemma \ref{lam0} (i) and (iii),
\begin{eqnarray*}
|J_{2n}(\beta_j^\star)|&\leq& cn^{-1/2}\sup|\sum_{i=1}^n \delta_i[x_{ij}-\widehat{\mathbb{D}}_{\beta_j^\star}
(\epsilon_i(\beta_j^\star))]|\sup \big| \frac{\int_{-\infty}^{\infty}ud\widehat{F}_{\beta_j^\star}(u)}{1-\widehat{F}_{\beta_j^\star}(u)}-\frac{\int_{-\infty}^{\infty}udF_{\beta_j^\star}(u)}{1-F_{\beta_j^\star}(u)}\big|\\
&=&O(n^{-1/2+3\varsigma+2\varepsilon})\; a.s.
\end{eqnarray*}

Since
$$\sum_{i=1}^n\{\mathbb{D}_{\beta_j^\star}(\epsilon_i(\beta_j^\star))-\widehat{\mathbb{D}}_{\beta_j^\star}
(\epsilon_i(\beta_j^\star))\}\epsilon_i(\beta_j^\star)\delta_i=\sum_{i=1}^n\int I(\zeta_i(\beta_j^\star)\ge u)\{\mathbb{D}_{\beta_j^\star}(u)-\widehat{\mathbb{D}}_{\beta_j^\star}
(u)\}u\frac{d\widehat{F}_{\beta_j^\star}(u)}{1-\widehat{F}_{\beta_j^\star}(u)},$$
then $J_{3n}$ can be written as
\begin{eqnarray*}
J_{3n}(\beta_j^\star)&=&n^{-1/2}\sum_{i=1}^n\int I(\zeta_i(\beta_j^\star)\ge u)\{\mathbb{D}_{\beta_j^\star}(u)-\widehat{\mathbb{D}}_{\beta_j^\star}
(u)\}d\mathcal{S}(u,\epsilon_i(\beta_j^\star)\mid F_{\beta_j^\star})\\
&=&n^{-1/2}\sum_{i=1}^n\{\mathbb{D}_{\beta_j^\star}(\epsilon_i(\beta_j^\star))-\widehat{\mathbb{D}}_{\beta_j^\star}
(\epsilon_i(\beta_j^\star))\}\epsilon_i(\beta_j^\star)\delta_i\\
&&-n^{-1/2}\sum_{i=1}^n\int I(\zeta_i(\beta_j^\star)\ge u)\{\mathbb{D}_{\beta_j^\star}(u)-\widehat{\mathbb{D}}_{\beta_j^\star}
(u)\}u\frac{dF_{\beta_j^\star}(u)}{1-F_{\beta_j^\star}(u)}\\
&=&n^{-1/2}\sum_{i=1}^n\int I(\zeta_i(\beta_j^\star)\ge u)\{\mathbb{D}_{\beta_j^\star}(u)-\widehat{\mathbb{D}}_{\beta_j^\star}
(u)\}u\frac{d\widehat{F}_{\beta_j^\star}(u)}{1-\widehat{F}_{\beta_j^\star}(u)}\\ &&-n^{-1/2}\sum_{i=1}^n\int I(\zeta_i(\beta_j^\star)\ge u)\{\mathbb{D}_{\beta_j^\star}(u)-\widehat{\mathbb{D}}_{\beta_j^\star}
(u)\}u\frac{dF_{\beta_j^\star}(u)}{1-F_{\beta_j^\star}(u)}\\
&=&n^{-1/2}\sum_{i=1}^n\int I(\zeta_i(\beta_j^\star)\ge u)\{\mathbb{D}_{\beta_j^\star}(u)-\widehat{\mathbb{D}}_{\beta_j^\star}
(u)\}u\big\{\frac{d\widehat{F}_{\beta_j^\star}(u)}{1-\widehat{F}_{\beta_j^\star}(u)}-\frac{dF_{\beta_j^\star}(u)}{1-F_{\beta_j^\star}(u)}\big\}\\
&\leq& n^{1/2}\sup_{u\leq \infty}\{|
\widehat{\mathbb{D}}_{\beta_j^\star}(u)-\mathbb{D}_{\beta_j^\star}(u)|\}
\sup\big| \frac{\int_{-\infty}^{\infty}ud\widehat{F}_{\beta_j^\star}(u)}{1-\widehat{F}_{\beta_j^\star}(u)}-\frac{\int_{-\infty}^{\infty}udF_{\beta_j^\star}(u)}{1-F_{\beta_j^\star}(u)}\big|
=O(n^{-1/2+3\varsigma+2\varepsilon})\;a.s.
\end{eqnarray*}
Hence, we complete the proof of Lemma \ref{lam1}.

\begin{lemma}\label{lam2}
$n^{1/2}\widetilde{\Psi}_n(\beta_j^\star)=n^{1/2}\widetilde{\Psi}_n(\beta_{0j}^\star)+V_n(\beta_j^\star-\beta_{0j}^\star)+o\{\max(n^{1/2},\boldsymbol{x}_j^{\T}\boldsymbol{x}_j)|\beta_j^\star-\beta_{0j}^\star|)\}$ a.s. for $|\beta_j^\star-\beta_{0j}^\star|\leq n^{-\psi}.$
\end{lemma}

\noindent {\large\bf Proof of Lemma \ref{lam2}}\\
Set
\begin{eqnarray*}
\widetilde{\Psi}_{n1}(a,\beta_j^\star)&=&
n^{-1/2}\sum_{i=1}^n\int I(\zeta_i(\beta_j^\star)\ge u)(x_{ij}-\mathbb{D}_{\beta_j^\star}
(u))W_{F_{\beta_j^\star}}(u)dF(u+aX_{ij}),\\
\widetilde{\Psi}_{n2}(a,\beta_j^\star)&=&n^{-1/2}\sum_{i=1}^n\int I(\zeta_i(\beta_j^\star)\ge u)(x_{ij}-\mathbb{D}_{\beta_j^\star}
(u))W_{F_{\beta_j^\star}}(u)\frac{\int_u^\infty dF(s+aX_{ij})}{1-F_{\beta_j^\star}(u)}dF_{\beta_j^\star}(u).
\end{eqnarray*}
Under the condition $\sup_i|x_{ij}|\leq c$, we have
$$\widetilde{\Psi}_{n}(\beta_j^\star)=\widetilde{\Psi}_{n1}(\beta_j^\star-\beta_{0j}^\star,\beta_j^\star)-
\widetilde{\Psi}_{n2}(\beta_j^\star-\beta_{0j}^\star,\beta_j^\star)+o(1),$$
for $|\beta_j^\star-\beta_{0j}^\star|\leq n^{-\psi}$.
Taking Taylor's expansion for $F_{\beta_j^\star}(u+aX_{ij})$ and $F_{\beta_j^\star}(s+aX_{ij})$, as $\beta_j^\star\rightarrow\beta_{0j}^\star$,
\begin{eqnarray*}
&&\widetilde{\Psi}_{n1}(\beta_j^\star-\beta_{0j}^\star,\beta_j^\star)-\widetilde{\Psi}_{n1}(0,\beta_j^\star)\\
&=&
n^{-1/2}\Big\{\sum_{i=1}^n\int I(\zeta_i(\beta_{0j}^\star)\ge u)x_{ij}(x_{ij}-\mathbb{D}_{\beta_{0j}^\star}
(u))W_{F}(u)df(u)\Big\}(\beta_j^\star-\beta_{0j}^\star)\\
&&+o(n^{-1/2}\boldsymbol{x}_j^{\T}\boldsymbol{x}_j|\beta_j^\star-\beta_{0j}^\star|)\\
&=&
n^{-1/2}\Big\{\sum_{i=1}^n\int I(\zeta_i(\beta_{0j}^\star)\ge u)x_{ij}(x_{ij}-\mathbb{D}_{\beta_{0j}^\star}
(u))W_{F}(u)\frac{f'(u)}{f(u)}dF(u)\Big\}(\beta_j^\star-\beta_{0j}^\star)\\
&&+o(n^{-1/2}\boldsymbol{x}_j^{\T}\boldsymbol{x}_j|\beta_j^\star-\beta_{0j}^\star|),
\end{eqnarray*}
and
\begin{eqnarray*}
&&\widetilde{\Psi}_{n2}(\beta_j^\star-\beta_{0j}^\star,\beta_j^\star)-\widetilde{\Psi}_{n2}(0,\beta_j^\star)\\
&=&n^{-1/2}\Big\{\sum_{i=1}^n\int I(\zeta_i(\beta_{0j}^\star)\ge u)x_{ij}(x_{ij}-\mathbb{D}_{\beta_{0j}^\star}
(u))W_{F}(u)\frac{\int_u^\infty df(s)}{1-F(u)}dF(u)\Big\}(\beta_j^\star-\beta_{0j}^\star)\\
&&+o(n^{-1/2}\boldsymbol{x}_j^{\T}\boldsymbol{x}_j|\beta_j^\star-\beta_{0j}^\star|)\\
&=&n^{-1/2}\Big\{\sum_{i=1}^n\int I(\zeta_i(\beta_{0j}^\star)\ge u)x_{ij}(x_{ij}-\mathbb{D}_{\beta_{0j}^\star}
(u))W_{F}(u)\frac{\int_u^\infty \frac{f'(s)}{f(s)}dF(s)}{1-F(u)}dF(u)\Big\}(\beta_j^\star-\beta_{0j}^\star)\\
&&+o(n^{-1/2}\boldsymbol{x}_j^{\T}\boldsymbol{x}_j|\beta_j^\star-\beta_{0j}^\star|).
\end{eqnarray*}
Therefore,
\begin{eqnarray*}
n^{1/2}\widetilde{\Psi}_{n}(\beta_j^\star)-n^{1/2}\widetilde{\Psi}_{n}(\beta_{0j}^\star)&=&
n^{1/2}\{\widetilde{\Psi}_{n1}(\beta_j^\star-\beta_{0j}^\star,\beta_j^\star)-\widetilde{\Psi}_{n2}(\beta_j^\star-\beta_{0j}^\star,\beta_j^\star)\}\\
&&-
n^{1/2}\{\widetilde{\Psi}_{n1}(0,\beta_{0j}^\star)-\widetilde{\Psi}_{n2}(0,\beta_{0j}^\star)\}+o(n^{1/2})\\
&=&V_n(\beta_j^\star-\beta_{0j}^\star)+o(\max\{n^{1/2},\boldsymbol{x}_j^{\T}\boldsymbol{x}_j|\beta_j^\star-\beta_{0j}^\star|\}),
\end{eqnarray*}
where
$$V_n=\sum_{i=1}^n\int I(\zeta_i(\beta_{0j}^\star)\ge u)x_{ij}(x_{ij}-\mathbb{D}_{\beta_{0j}^\star}
(u))W_{F}(u)W_{F}\big(u,f'/f)dF(u),
$$
where
$$W_{F}\big(u,h)
=h(u)-\frac{\int_u^\infty h(s)dF(s)}{1-F(u)}.$$

\vspace{.2in}
\noindent {\large\bf Proof of Theorem 1}\\
\vspace{.1in}
Lemma \ref{lam1} is equivalent to
\begin{eqnarray}\label{pthm1}
\sup_{|\beta_j^\star|\leq\rho}|\Psi_n(\beta_j^\star)-\widetilde{\Psi}_n(\beta_j^\star)|=o(n^{-1/2+4\varsigma})\;a.s.
\end{eqnarray}
Under the condition $\lim_{n\rightarrow\infty}n^{1/2-4\varsigma}\Big\{
\inf_{|\beta_j^\star|\leq\rho,|\beta_j^\star-\beta_{0j}^\star|\ge n^{-\psi}}|\widetilde{\Psi}_n(\beta_j^\star)|\Big\}=\infty$ and (\ref{pthm1}),
\begin{eqnarray*}
P\{\Psi_n(\beta_j^\star)\; \mbox{has \;a \;zero-crossing\; on}\; |\beta_j^\star-\beta_{0j}^\star|\ge n^{-\psi}\;{\rm and}\;|\beta_j^\star|\leq\rho\;{\rm for\;large}\; n\}=0.
\end{eqnarray*}
Since $\widehat{\beta}_j^\star$, $\Psi_n(\widehat{\beta}_j^\star)=0$, then by Lemma \ref{lam2} and conditions $\boldsymbol{x}_j^{\T}\boldsymbol{x}_j\leq n$ and $4\varsigma+\gamma>1$ with $\frac{1}{8}\leq\varsigma<1$,
we have \begin{eqnarray}\label{pthm3}
\sup_{|\widehat{\beta}_j^\star-\beta_{0j}^\star|\leq n^{-\psi}}|\widetilde{\Psi}_n(\beta_{0j}^\star)+n^{-1/2}V_n(\widehat{\beta}_j^\star-\beta_{0j}^\star)|=o(n^{-1/2+4\varsigma})\;a.s.
\end{eqnarray}
 Since $E\{\widetilde{\Psi}_n(\beta_{0j}^\star)\}=0$, we have $|n^{1/2}\widetilde{\Psi}_n(\beta_{0j}^\star)|=O(n^{1/2+\varepsilon})\;a.s..$
Therefore, under $|1/V_n|\leq cn^{-1}$, Condition (C3) (ii) and $\xi t+\sqrt{2t}\gg n^{4\varsigma-1}$ for any constant $t$, we utilize the Bernstein's inequality and have
\begin{eqnarray*}
P\Big(|\widehat{\beta}_j^\star-\beta_{0j}^\star|\ge \xi t+\sqrt{2t}\Big)\leq
P\Big(|n^{-1/2}\widetilde{\Psi}_n(\beta_{0j}^\star)|\ge \xi t+\sqrt{2t}\Big)\leq\exp(-nt).
\end{eqnarray*}
Next we prove Theorem 1 (ii). (\ref{Psin}) implies that $\beta_{0j}^\star$ satisfies the marginal Buckley-James score function
\begin{eqnarray}\label{estim}
E\{(X_{j}-\bar{X}_j)(X^{\T}\boldsymbol{\beta}_0-X_{j}\beta_{0j}^\star)\}=0.
\end{eqnarray}
Under additional conditions $Ex_j=0$, $Ex_j^2=1$ and ${\rm cov}(X^{\T}\boldsymbol{\beta}_0,x_j)\ge \xi t+\sqrt{2t}$ for any constant $t$ and $j=1,\cdots,p,$ we have
$$\min_{j\in\mathcal{D}}|\beta_{0j}^\star|\ge \xi t+\sqrt{2t}.$$
Note that on the event
$A_n=\{\max_{j \in \mathcal{D}}|\widehat{\beta}_{j}^\star-\beta_{0j}^\star|\leq (\xi t+\sqrt{2t})/2\},$
 we have $$|\widehat{\beta}_{j}^\star|\ge( \xi t+\sqrt{2t})/2\;\; {\rm for}\; {\rm all}\; j\in\mathcal{D}.$$
Hence, by the choice of $\gamma_n$ , we have $\mathcal{D}\subset\widehat{\mathcal{D}}$. The result now follows from a
simple union bound
$$P(A_n^c)\leq d\exp(-\frac{nt}{4}).$$
We next prove the third part of Theorem 1. Denote $\Sigma=E(XX^{\T})$, the marginal Buckley-James score function (\ref{estim}) implies that
 $$\|\boldsymbol{\beta}_0^\star\|^2=||\Sigma\boldsymbol{\beta}_0||^2=O\{\lambda_{\max}(\Sigma)\},$$
 where $\boldsymbol{\beta}_0^\star=(\beta_{01}^\star,\cdots,\beta_{0p}^\star)^{\T} $, then the number of $\{j:|\beta_j^\star|>(\xi t+\sqrt{2t})/2\}$ cannot exceed the number of  $||\Sigma\boldsymbol{\beta}_0||/(\xi t+\sqrt{2t})^2$. Thus on the set $A_n$,
the number of $\{j:|\widehat{\beta}_j^\star|> (\xi t+\sqrt{2t})\}$
cannot exceed the number of $\{j:|\beta_j^\star|> (\xi t+\sqrt{2t})\}/2$, which is bounded by $O\{\lambda_{\max}(\Sigma)/(\xi t+\sqrt{2t})^2\}$. Then we have
$$P\Big(|\widehat{\mathcal{D}}|\leq O\{\lambda_{\max}(\Sigma)/(\xi t+\sqrt{2t})^2\}\Big)\ge P(A_n)\ge 1- d\exp(-\frac{nt}{4})$$
which completes the proof of the theorem.

\begin{lemma}\label{Lem1} Under the conditions (i)The noise vector $\boldsymbol{\epsilon}$ has sub-Gaussian tails such that $P(|\tau^{\T}\boldsymbol{\epsilon}|<\|\tau\|x) \geq
    1-2\exp(-c_{1} x^2)$ for any vector $\tau \in \mathbb{R}^{n}$, $0<c_{1}<\infty$ and $x>0$, and $E (\epsilon_{i}^4)<c_2<\infty$ for $i=1,\cdots, n$.
 (ii) The eigenvalue of $\gamma(\boldsymbol{X}^{\T}\boldsymbol{X})$ is bounded.  $\sup_{i}\|\boldsymbol{x}_{i}\| \leq c_3\sqrt{q}$, for some positive constants $c_{3}$. 

$Y_{ni}=E(Y_i\mid X_i)+\mathcal{S}(\zeta_i(\boldsymbol{\beta}),\epsilon_i(\boldsymbol{\beta})\mid F_{\boldsymbol{\beta}})-E\{\mathcal{S}(\zeta_i(\boldsymbol{\beta}),\epsilon_i(\boldsymbol{\beta})\mid F_{\boldsymbol{\beta}})\}+O_p(\sqrt{q/n})$.
\end{lemma}
\noindent  \emph{\bf Proof of Lemma \ref{Lem1}}.
Based on $Y_{ni}-E(Y_i\mid X_i)=Y_i(\widehat{\boldsymbol{\beta}},\widehat{F})-
Y_i(\boldsymbol{\beta},F)+Y_i(\boldsymbol{\beta},F)-E(Y_i\mid X_i)=I_1+I_2,$
and
$$|I_1|\leq\big| \frac{\int_t^{\infty}sd\widehat{F}_{\widehat{\boldsymbol{\beta}}}(s)}{1
-\widehat{F}_{\widehat{\boldsymbol{\beta}}}(t)}-\frac{\int_t^{\infty}sd\widehat{F}_{\boldsymbol{\beta}}(s)}{1
-\widehat{F}_{\boldsymbol{\beta}}(t)}\big|+\big| \frac{\int_t^{\infty}sd\widehat{F}_{\boldsymbol{\beta}}(s)}{1
-\widehat{F}_{\boldsymbol{\beta}}(t)}-\frac{\int_t^{\infty}sdF_{\boldsymbol{\beta}}(s)}{1
-F_{\boldsymbol{\beta}}(t)}\big|=|I_{11}|+|I_{12}|,$$
$|I_{11}|$ can be bounded by high-dimensional consistency rate given by $||\widehat{\boldsymbol{\beta}}-\boldsymbol{\beta}||=||\widehat{\boldsymbol{\beta}}_1-\boldsymbol{\beta}_1||=O_p(\sqrt{q/n}),$
 (Wang et al., 2008).
For $|I_{12}|$, we utilized Lemma 2 of Lai and Ying (1991), i.e.,
\begin{eqnarray*}
&&\sup\Big\{\big| \frac{\int_t^{\infty}sd\widehat{F}_{\boldsymbol{\beta}}(s)}{1
-\widehat{F}_{\boldsymbol{\beta}}(t)}-\frac{\int_t^{\infty}sdF_{\boldsymbol{\beta}}(s)}{1
-F_{\boldsymbol{\beta}}(t)}\big|:\sup_j|\beta_j|\leq \kappa,t\leq s\leq \infty,
\sum_{i=1}^nI(\upsilon_i(\boldsymbol{\beta})\ge s)\ge \frac{cn}{2}\Big\}\\
&&=O(n^{-1/2+\varepsilon})\; a.s.
\end{eqnarray*}  for every
$0\leq\varsigma<1$ and $\varepsilon>0$. Note that $Y_i(\boldsymbol{\beta},F)-E(Y_i\mid X_i)=\mathcal{S}(\zeta_i(\boldsymbol{\beta}),\epsilon_i(\boldsymbol{\beta})\mid F_{\boldsymbol{\beta}})-E\{\mathcal{S}(\zeta_i(\boldsymbol{\beta}),\epsilon_i(\boldsymbol{\beta})\mid F_{\boldsymbol{\beta}})\}$. Then we have proved Lemma \ref{Lem1}.

\begin{lemma}\label{Lem2}
Under Condition C2 (i), there exists a positive constant $B$ such that
\begin{itemize}
\item[(i)] $\lambda_{\rm max}(H_k-\tilde{H}_k)\leq B\times \frac{p_k}{n}$;

\item[(ii)] tr$[(\tilde{H}_k-H_k)^{\T}(\tilde{H}_k-H_k)]\leq B^2\times \frac{p_k^2}{n}$;

\item[(iii)] tr$[(\tilde{H}_k-H_k)^{\T}(\tilde{H}_k-H_k)]^2\leq B\times \frac{p_k^4}{n^3}$;

\item[(iv)] $\lambda_{\rm max}(\tilde{H}_k) \leq 1+B\times \frac{p_k}{n}$;

\item[(v)] tr$(\tilde{H}_k\tilde{H}_k^{\T})\leq B\times p_k$.
\end{itemize}
\end{lemma}
\noindent  \emph{\bf Proof of Lemma \ref{Lem2}}.
We can refer to Lemma 3.1 of Ando and Li (2014) to prove Lemma \ref{Lem2}.

\noindent  \emph{\bf Proof of Theorem 2}.
Let $\boldsymbol{S}$ be
an $n$-vector
with the $i$th component $\mathcal{S}(\zeta_i(\boldsymbol{\beta}),\epsilon_i(\boldsymbol{\beta})\mid F_{\boldsymbol{\beta}})-E\{\mathcal{S}(\zeta_i(\boldsymbol{\beta}),\epsilon_i(\boldsymbol{\beta})\mid F_{\boldsymbol{\beta}})\}$. $\Sigma=E(\boldsymbol{S}^{\T}\boldsymbol{S})$ is a diagonal matrix whose diagonal element is  $\sigma^2_S=E[\mathcal{S}(\zeta_i(\boldsymbol{\beta}),\epsilon_i(\boldsymbol{\beta})\mid F_{\boldsymbol{\beta}})]^2$.
Denote $\tilde{L}(\boldsymbol{\omega})=(\boldsymbol{\mu}-\tilde{\boldsymbol{\mu}}(\boldsymbol{\omega}))^{\T}(\boldsymbol{\mu}-\tilde{\boldsymbol{\mu}}(\boldsymbol{\omega}))$ with $\tilde{\boldsymbol{\mu}}(\boldsymbol{\omega})=\tilde{H}(\boldsymbol{\omega})\boldsymbol{Y}_n$, and $\tilde{A}(\boldsymbol{\omega})=I-\tilde{H}(\boldsymbol{\omega})$. By Lemma \ref{Lem1}, we have
\[
\begin{array}{llll}
\mathcal{M}(\boldsymbol{\omega})
&=&\{||\boldsymbol{S}||^2+\tilde{L}(\boldsymbol{\omega})+2<\boldsymbol{S}, \boldsymbol{\mu}-\tilde{H}(\boldsymbol{\omega})\boldsymbol{Y}_n>\}+O_p(K\sqrt{q/n})\\
&=&||\boldsymbol{S}||^2+L(\boldsymbol{\omega})\{\frac{\tilde{L}(\boldsymbol{\omega})}{L(\boldsymbol{\omega})}+\frac{2<\boldsymbol{S}, \boldsymbol{\mu}-\tilde{H}(\boldsymbol{\omega})\boldsymbol{Y}_n>/R(\boldsymbol{\omega})}{L(\boldsymbol{\omega})/R(\boldsymbol{\omega})}\}+O_p(K\sqrt{q/n}).
 \end{array}
\]
Note that the acquisition  of  $\widehat{\boldsymbol{\omega}}$  is equivalent to minimize $\tilde{ \mathcal{M}}(\boldsymbol{\omega})=\mathcal{M}(\boldsymbol{\omega})-||\boldsymbol{S}||^2$ over $\boldsymbol{\omega}\in\mathcal{V}$. Under Condition (C2) (v), $\frac{L(\widehat{\boldsymbol{\omega}})}{\inf\limits_{\omega\in\mathcal{V}}L(\boldsymbol{\omega})}\rightarrow 1$ if we can show that
\begin{eqnarray}{\label{DWcv1}}
\sup\limits_{\boldsymbol{\omega}\in\mathcal{V}}|\tilde{L}(\boldsymbol{\omega})/L(\boldsymbol{\omega})-1|\rightarrow 0,\\\label{DWcv2}
 \sup\limits_{\boldsymbol{\omega}\in\mathcal{V}}|<\boldsymbol{S}, \boldsymbol{\mu}-\tilde{H}(\boldsymbol{\omega})\boldsymbol{Y}_n>|/R(\boldsymbol{\omega})\rightarrow 0,\\\label{DWcv3}
\sup\limits_{\boldsymbol{\omega}\in\mathcal{V}}L(\boldsymbol{\omega})/R(\boldsymbol{\omega})\rightarrow 1.
\end{eqnarray}
By Canchy-Schwartz inequality,
\[
\begin{array}{llll}
|\tilde{L}(\boldsymbol{\omega})-L(\boldsymbol{\omega})|&=&\mid||\{\tilde{H}(\boldsymbol{\omega})-H(\boldsymbol{\omega})\}\boldsymbol{Y}_n||^2-2<\boldsymbol{\mu}-
H(\boldsymbol{\omega})\boldsymbol{Y}_n,\{\tilde{H}(\boldsymbol{\omega})-H(\boldsymbol{\omega})\}\boldsymbol{Y}_n>\mid\\
&\leq&||\{\tilde{H}(\boldsymbol{\omega})-H(\boldsymbol{\omega})\}\boldsymbol{Y}_n||^2
+2\sqrt{L(\boldsymbol{\omega})}||\{\tilde{H}(\boldsymbol{\omega})-H(\boldsymbol{\omega})\}\boldsymbol{Y}_n||.
 \end{array}
\]
Note that
\[
\begin{array}{llll}
&&||\{\tilde{H}(\boldsymbol{\omega})-H(\boldsymbol{\omega})\}\boldsymbol{Y}_n||^2\\
&\leq& \{\sum\limits_{k=1}^K\omega_k||(\tilde{H}_k-H_k)\boldsymbol{\mu}||+\sum\limits_{k=1}^K\omega_k||(\tilde{H}_k-H_k)\boldsymbol{S}||\}^2\\
&\leq& 2K^2\{\max\limits_k||(\tilde{H}_k-H_k)\boldsymbol{\mu}||^2+\max\limits_k||(\tilde{H}_k-H_k)\boldsymbol{S}||^2\}.\\
\end{array}
\]
The proof of the first part follows from Ando and Li (2014). It is sufficient to show that
\begin{eqnarray}\label{DWcv12} K^2\max\limits_k||(\tilde{H}_k-H_k)\boldsymbol{S}||^2/\xi_n\rightarrow 0,
 \end{eqnarray}
i.e., it is sufficient to verify 
\begin{eqnarray}{\label{DWcv120}}
&K^2\max\limits_kE||(\tilde{H}_k-H_k)\boldsymbol{S}||^2/\xi_n\rightarrow 0, \\ {\label{DWcv121}}
&{\rm and}\quad \sum_{k=1}^KP\{K^2|||(\tilde{H}_k-H_k)\boldsymbol{S}||^2-E||(\tilde{H}_k-H_k)\boldsymbol{S}||^2|/\xi_n>\tau\}\rightarrow 0.
\end{eqnarray}
 For (\ref{DWcv120}), utilizing the inequality $tr(ABA^{\T})\leq \lambda_{\max}(B)tr(AA^{\T})$ and Lemma \ref{Lem2} (ii), we have
 \[
\begin{array}{llll}
E||(\tilde{H}_k-H_k)\boldsymbol{S}||^2&=&tr\{(\tilde{H}_k-H_k)\Sigma(\tilde{H}_k-H_k)^{\T}\}\\
&\leq& \lambda_{\max}(\Sigma)tr\{(\tilde{H}_k-H_k)(\tilde{H}_k-H_k)^{\T}\}
\leq B^2\lambda_{\max}(\Sigma)\frac{p_k^2}{n}.
\end{array}
\]
So (\ref{DWcv120}) holds. We can show (\ref{DWcv121}) in the following way:
\[
\begin{array}{llll}
\sum_{k=1}^KP\{K^{4\kappa}\mid||(\tilde{H}_k-H_k)\boldsymbol{S}||^2/\xi_n-E||(\tilde{H}_k-H_k)\boldsymbol{S}||^2\mid^{2\kappa}/\xi_n^{2\kappa}>\tau^{2\kappa}\}\\
\leq\sum\limits_{k=1}^K\frac{K^{4\kappa}E\{||(\tilde{H}_k-H_k)\boldsymbol{S}||^2/\xi_n-E||(\tilde{H}_k-H_k)\boldsymbol{S}||^2\}^{2\kappa}}{\xi_n^{2\kappa}\tau^{2\kappa}}\\ 
\leq B_4\times B^\kappa \frac{K^{4\kappa}}{\xi_n^{2\kappa}\tau^{2\kappa}}\sum\limits_{k=1}^K(p_k^4/n^3)^{\kappa}
\leq B_4\times B^\kappa \times B_2^{4\kappa} \frac{K^{4\kappa+1}}{\xi_n^{2\kappa}\tau^{2\kappa}}\rightarrow 0.
\end{array}
\]
Then it yields  (\ref{DWcv1}).

Note that  (\ref{DWcv2}) holds if  we can show that
\begin{eqnarray}{\label{DWcv22}}
\sup\limits_{\boldsymbol{\omega}\in\mathcal{V}}|<\boldsymbol{S}, \boldsymbol{\mu}>|/R(\boldsymbol{\omega})\rightarrow 0,\\{\label{DWcv23}}
\sup\limits_{\boldsymbol{\omega}\in\mathcal{V}}|<\boldsymbol{S}, \tilde{H}(\boldsymbol{\omega})\boldsymbol{\mu}>|/R(\boldsymbol{\omega})\rightarrow 0,\\{\label{DWcv24}}
\sup\limits_{\boldsymbol{\omega}\in\mathcal{V}}|<\boldsymbol{S}, \tilde{H}(\boldsymbol{\omega})\boldsymbol{S}>|/R(\boldsymbol{\omega})\rightarrow 0.
\end{eqnarray}
Using the  Bonferroni's inequality,
we obtain that for any $\tau$,
\[
\begin{array}{llll}
P\{\sup\limits_{\boldsymbol{\omega}\in\mathcal{V}}|<\boldsymbol{S}, \boldsymbol{\mu}>|/R(\boldsymbol{\omega})>\tau\}
\leq P\{(\boldsymbol{S}^{\T}\boldsymbol{\mu})>\tau\xi_n\}
\leq \frac{\{E\boldsymbol{\mu}^{\T}\boldsymbol{S}\boldsymbol{S}^{\T}\boldsymbol{\mu}\}^{\kappa}}{\tau^{2\kappa}\xi_n^{2\kappa}}
\leq \sigma^{2\kappa}_S \frac{||\boldsymbol{\mu}||^{2\kappa}}{\tau^{2\kappa}\xi_n^{2\kappa}} \rightarrow 0,
\end{array}
\]
which yields (\ref{DWcv22}).

 For (\ref{DWcv23}), under Condition (C3) (iii), we have
 \[
\begin{array}{llll}
P\{\sup\limits_{\boldsymbol{\omega}\in\mathcal{V}}|<\boldsymbol{S}, \tilde{H}(\boldsymbol{\omega})\boldsymbol{\mu}>|/R(\boldsymbol{\omega})>\tau\}
\leq P\{K\max\limits_k|<\boldsymbol{S}, \tilde{H}_k(\boldsymbol{\omega})\boldsymbol{\mu}>|/R(\boldsymbol{\omega})>\tau\}\\
\leq\sum_{k=1}^KP\{K(\boldsymbol{S}^{\T}\tilde{H}_k\boldsymbol{\mu})>\tau\xi_n\}
\leq K^{2\kappa+1}\frac{(E\boldsymbol{\mu}\tilde{H}_k^{\T}\boldsymbol{S}\boldsymbol{S}^{\T}\tilde{H}_k\boldsymbol{\mu} )^{\kappa}}{\tau^{2\kappa}\xi_n^{2\kappa}}
\leq \sigma^{2\kappa}_S\lambda_{\max}^\kappa(\tilde{H}_k^{\T}\tilde{H}_k)\frac{K^{2\kappa+1}||\boldsymbol{\mu}||^{2\kappa}}{\tau^{2\kappa}\xi_n^{2\kappa}} \rightarrow 0.
\end{array}
\]

For (\ref{DWcv24}),  we have
 \[
\begin{array}{llll}
P\{\sup\limits_{\boldsymbol{\omega}\in\mathcal{V}}|<\boldsymbol{S}, \tilde{H}(\boldsymbol{\omega})\boldsymbol{S}>|/R(\boldsymbol{\omega})>\tau\}
\leq P\{K\max\limits_k|<\boldsymbol{S}, \tilde{H}_k(\boldsymbol{\omega})\boldsymbol{S}>|/R(\boldsymbol{\omega})>\tau\}\\
\leq\sum_{k=1}^KP\{K(\boldsymbol{S}^{\T}\tilde{H}_k\boldsymbol{S})>\tau\xi_n\}
\leq K^{2\kappa+1}\frac{(E\boldsymbol{S}\tilde{H}_k^{\T}\boldsymbol{S}\boldsymbol{S}^{\T}\tilde{H}_k\boldsymbol{S}^{\T} )^{\kappa}}{\tau^{2\kappa}\xi_n^{2\kappa}}
\leq \sigma^{4\kappa}_S\lambda_{\max}^\kappa(\tilde{H}_k^{\T}\tilde{H}_k)\frac{K^{2\kappa+1}}{\tau^{2\kappa}\xi_n^{2\kappa}} \rightarrow 0.
\end{array}
\]

Note that (\ref{DWcv3}) is equivalent to
 \[
\begin{array}{llll}
\sup\limits_{\boldsymbol{\omega}\in\mathcal{V}}|\frac{||H(\boldsymbol{\omega})\boldsymbol{S}||^2-tr(H(\boldsymbol{\omega})\Sigma H^{\T}(\boldsymbol{\omega}))-2<A(\boldsymbol{\omega})\boldsymbol{\mu},H(\boldsymbol{\omega})\boldsymbol{S}>}{R(\boldsymbol{\omega})}|\rightarrow 0,
\end{array}
\]
where $A(\boldsymbol{\omega})=I-H(\boldsymbol{\omega})$.
Under Condition (C3) (iii),  it suffices to show that
\begin{equation}{\label{DWcv31}}
\sup\limits_{\boldsymbol{\omega}\in\mathcal{V}}|\frac{||H(\boldsymbol{\omega})\boldsymbol{S}||^2-tr(H(\boldsymbol{\omega})\Sigma H^{\T}(\boldsymbol{\omega}))}{R(\boldsymbol{\omega})}|\rightarrow 0
\end{equation}
and
\begin{equation}{\label{DWcv32}}
\sup\limits_{\boldsymbol{\omega}\in\mathcal{V}}|\frac{<A(\boldsymbol{\omega})\boldsymbol{\mu},H(\boldsymbol{\omega})\boldsymbol{S}>}{R(\boldsymbol{\omega})}|\rightarrow 0.
\end{equation}
Now we prove (\ref{DWcv31}) and (\ref{DWcv32}). For any $\tau>0$, we have
 \[
\begin{array}{llll}
&P(\sup\limits_{\boldsymbol{\omega}\in\mathcal{V}}|\frac{||H(\boldsymbol{\omega})\boldsymbol{S}||^2-\sigma^2_S tr(H(\boldsymbol{\omega}) H^{\T}(\boldsymbol{\omega}))}{R(\boldsymbol{\omega})}|>\tau)\\
&\leq P(\sup\limits_{\boldsymbol{\omega}\in\mathcal{V}}\sum\limits_{k=1}^K\sum\limits_{j=1}^K|\boldsymbol{S}^{\T}H_kH_j\boldsymbol{S}-\sigma^2_S tr(H_k H_j)|>\tau\xi_n)\\
&\leq\sum\limits_{k=1}^K\sum\limits_{j=1}^KP(\max\limits_{k}\max\limits_{j}|\boldsymbol{S}^{\T}H_kH_j\boldsymbol{S}-\sigma^2_S tr(H_k H_j)|>\tau\xi_n/K^2)\\
&\leq B\frac{K^{4\kappa}}{\tau^{2\kappa}\xi_n^{\kappa}}\sum\limits_{k=1}^K\sum\limits_{j=1}^K \{tr(H^2_kH_j^2)\}^\kappa\leq B\frac{K^{4\kappa+2}n^\kappa}{\tau^{2\kappa}\xi_n^{2\kappa}}\rightarrow 0,
\end{array}
\]
 where Condition (C3) (iii) ensures that the last term converges to 0. Similarly, for any $\tau>0$,
 \[
\begin{array}{llll}
&P(\sup\limits_{\boldsymbol{\omega}\in\mathcal{V}}|\frac{<A(\boldsymbol{\omega})\boldsymbol{\mu},H(\boldsymbol{\omega})\boldsymbol{S}>}{R(\boldsymbol{\omega})}|>\tau)\\
&\leq P(K^2\max\limits_{k}\max\limits_{j}|\boldsymbol{\mu}^{\T}(I-H_k)H_j\boldsymbol{S}|>\tau\xi_n)\\
&\leq B \frac{K^{4\kappa}}{\tau^{2\kappa}\xi_n^{2\kappa}}\sum\limits_{k=1}^K\sum\limits_{j=1}^K||H_j(I-H_k)\boldsymbol{\mu}||^{2\kappa}
\leq B\frac{K^{4\kappa+2}||\boldsymbol{\mu}||^{2\kappa}}{\tau^{2\kappa}\xi_n^{2\kappa}}\rightarrow 0.
\end{array}
\]
Thus   (\ref{DWcv3}) holds.

\vskip .65cm
\noindent


\end{document}